\documentclass[preprint,12pt]{elsarticle}

\usepackage{amssymb}
\usepackage{amsmath}
\usepackage{parskip}
\usepackage{multirow}

\usepackage{pdfpages}
\usepackage{graphicx}
\usepackage{graphics}
\usepackage {framed}
\usepackage {amsxtra}
\usepackage {enumerate}
\usepackage{commath}
\usepackage[margin=1 in]{geometry}
\usepackage{float}
\usepackage[caption = false]{subfig}

\journal{arXiv}

\begin{document}

\begin{frontmatter}



\title{Nonparametrically estimating dynamic bivariate correlation using visibility graph algorithm}


\author{Aparna John$^{a, *,}$\footnote{$^{*} =$ contributed equally} , Toshikazu Ikuta$^{b, *}$, Janina D Ferbinteanu$^{c,
**,}$\footnote{$^{**} =$ contributed equally},
 Majnu John$^{d,}$$^{e,}$$^{f, **,}$\footnote{Corresponding author: $350$ Community Drive, Manhasset, NY 11030.
 e-mail: {\sf mjohn5@northwell.edu, majnu.john@hofstra.edu}, Phone: +01\,718\,470\,8221, Fax: +01\,718\,343\,1659}}

\address{$^{a}$Department of Electrical and Computer Engineering,\\ The University of Texas at San Antonio,\\San Antonio, TX.}
\address{$^{b}$Department of Communication Sciences and Disorders, School of Applied Sciences, \\University of Mississippi,
\\University, MS.}
\address{$^{c}$Departments of Physiology and Pharmacology and of Neurology, \\State University of New York Downstate Medical
Center, \\Brooklyn, NY.}
\address{$^{d}$Center for Psychiatric Neuroscience, \\Feinstein Institute of Medical Research, \\Manhasset, NY.}
\address{$^{e}$Division of Psychiatry Research, The Zucker Hillside Hospital,\\Northwell Health System,  \\Glen Oaks, NY.}
\address{$^{f}$Department of Mathematics, \\Hofstra University, \\Hempstead, NY.}

\begin{abstract}

Dynamic conditional correlation (DCC) is a method that estimates the correlation between two time series across time.  Although
used primarily in finance so far, DCC has been proposed recently as a model-based estimation method for quantifying functional connectivity during fMRI experiments. DCC could also be used to estimate the dynamic correlation between other types of time series such as local field potentials (LFP's) or spike trains recorded from distinct brain areas.
DCC  has very nice properties compared to other existing methods, but its applications for neuroscience are currently limited because of non-optimal performance in the presence of outliers. To address this issue,
we developed a robust nonparametric version of DCC, based on an adaptation of the weighted visibility graph algorithm which converts a time series into a weighted graph. The modified DCC demonstrated better performance in the analysis of empirical data sets: one fMRI data set collected from a  human subject performing a Stroop task; and one LFP data set recorded from an awake rat in resting state. Nonparametric DCC has the potential of
enlarging the spectrum of analytical tools designed to assess the dynamic coupling and uncoupling of activity among brain areas.


\end{abstract}


\end{frontmatter}





\noindent

\section*{Introduction}

Estimation methods for dynamic bivariate correlation in resting state fMRI have been gaining increasing attention in the neuroimaging community.
Dynamic correlation estimation is of interest because many recent studies have identified dynamic changes in functional connectivity during the
course of an fMRI experiment [1-7], especially during resting state. Estimating dynamic correlation is of interest in
neuroscience in general. For example, dynamic correlation between local field potential time series obtained from different brain regions could be
used to explore how certain brain regions work in tandem during certain specific behaviors. When identifying such changes, it is of importance to ensure that the dynamic
shifts in the correlations observed are not due to spurious fluctuations inherent to the estimation method used. This fact underlines the need for
assessing the accuracy of existing estimation methods and for developing accurate new methods.

Recently Lindquist and co-authors [8] published an expository paper in which they recommended using the dynamic conditional correlation (DCC) method
[9,10] as the best approach for estimating the temporal evolution of the correlation between two time series. Two other methods that Lindquist and
co-authors considered were the commonly used sliding window technique and the exponential weighted moving average (EWMA) technique. Using a
simulation study and real-data examples, Lindquist \textit{et al.} [8] illustrated that DCC performed far better when compared to the other two
methods.

DCC is widely used in scientific fields other than neuroscience (e.g. finance), but it has some often overlooked shortcomings. DCC is a parametric
method, in which the parameters involved are estimated using a 2-stage maximum likelihood (ML) method. The optimization task in the ML method is
conducted using numerical iteration methods for which initial values for the parameters have to be provided. Numerical convergence of these methods
typically does not depend on the starting points. However, while working on the real-data examples and simulations for this paper, we encountered a
few scenarios where convergence (or rather non-convergence) could pose a problem, at least for the iterative methods used in two different R packages
available for DCC: \textit{ccgarch} and \textit{rmgarch}.

A distinct and more important issue with the parameter framework of DCC is its sensitivity to extreme values, which do occur in time series measurements and can be ignored only if such values are an anomaly that does not make sense with the underlying scientific framework. A good example comes from financial time series. Any financial time series based on US markets (e.g. S$\&$P index) had a few extreme values during the 2008 market crash. Dynamic correlations considered for such a pair of time series, if including the values from 2008, cannot ignore those extreme values, because they are real. Time series examples with normally occurring extreme values in neuroimaging and neuroscience are not uncommon either, as seen, for example, in the LFP time series data, which is one of the data sets analyzed in this paper.

To address these problems, we propose a novel nonparametric approach for estimating dynamic correlation estimation which we refer to as
nonparametric weighted visibility graph algorithm (nonparametric WVGA). Our method is based on an adaptation of the visibility graph algorithm (VGA),
introduced by physicists, Lacasa and co-authors [11]. VGA's general approach is to convert time series into mathematical graphs, a procedure that
does not involve any parameters. Therefore, our new method is nonparametric in nature and robust to extreme values. We illustrate the
utility of this method via simulations and real data examples.

\section*{Methods}

For the current work, we considered three methods for estimating dynamic correlation (Figure 1). The first two methods - the SW and DCC - are
currently existing methods and here we followed closely the exposition and notation used in Lindquist \textit{et al} [8]. The third method based on
weighted visibility graph algorithm is our new approach to the dynamic-correlation-estimation problem.

We illustrate the methods using two time series generated as in Lindquist \textit{et al}'s [8]
simulations study $\#1$. Random data were generated for each time point
using a mean-zero bivariate normal distribution, with correlation matrix set to
\[  \left[ \begin{array}{cc}
 \sqrt{2} &  0              \\
  0    & \sqrt{3}           \end{array} \right].  \]
A bivariate time series generated using the above mechanism with $T = 300$ time points is shown in the top panel in Figure 1. The dynamic
correlation was \textit{a priori} set to zero  in this simulated example. That is, the underlying correlation of this simulated bivariate time series equals zero at all points in time.

\begin{figure}[H]
\begin{center}
\includegraphics[height=3in,width=7in,angle=0]{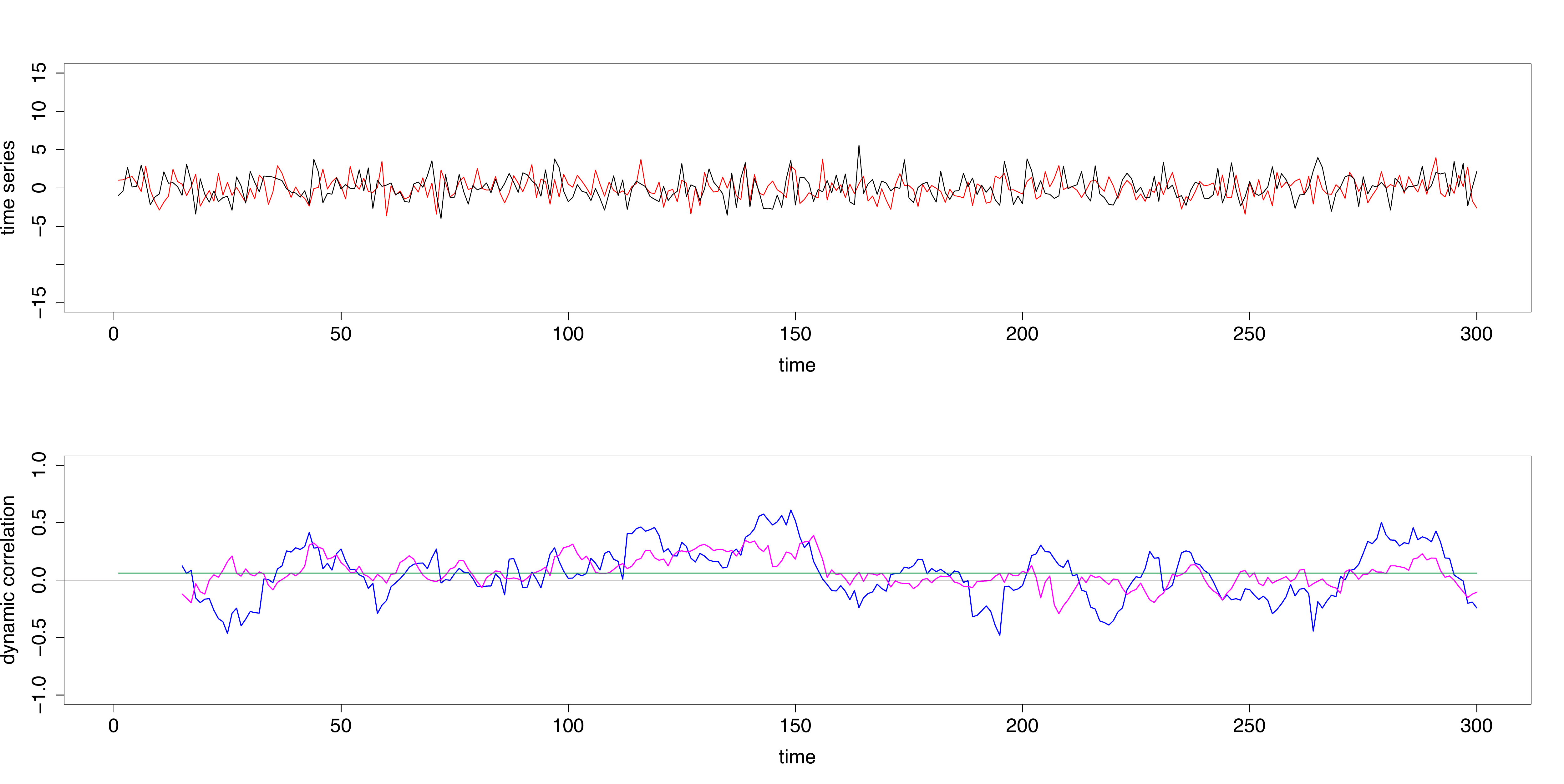}
\caption{Top panel: The two time series used for the first illustrative simulated example; data were generated
from a bivariate normal distribution. Bottom Panel: Dynamic correlations estimated for the
above two time series using 3 different methods: method $\#$1, SW (blue); method $\#$2, DCC (green);
and method $\#$3, nonparametric WVGA (magenta). The black horizontal line in the center
through zero represents the underlying true dynamic correlation for this example.}
\end{center}
\end{figure}

\subsection*{Method $\#$ 1 - Sliding-Window (SW) technique} SW technique [1,2,4] is easy to explain and
implement. We chose a window-frame of size ``$ws$" = 15 for the example considered in
figure 1. At the first iteration, we used the window-frame to frame the first
($ws =$) 15 time points of the bivariate time series, and
find the correlation between the values of the two series within this window-frame. Then we repeat the process by sliding the window to the right, one unit at a time. Thus, in general, at iteration $i$, the right-end of the SW was placed at the $(i - 1 + ws)^{th}$ time point of both time series, and the correlation between the two time series within the window-frame was calculated.  This series of correlations forms the blue curve plotted in the bottom-panel of Figure 1.

To make it clearer, the $1^{st}$ correlation value is the correlation between two vectors each of length 15, each containing the values corresponding to time points between 1 and 15 (end points included) of the $1^{st}$ and the $2^{nd}$ time series, respectively. To obtain the second correlation, we moved the window one unit to the right, so that for this second iteration the right-end of the window-frame is at time-point 16, and the left end is at time-point 2. The $2^{nd}$ correlation value will be the correlation between two vectors each of length 15, each containing the values corresponding to time points between 2 and 16 (end points included) of the $1^{st}$
and the $2^{nd}$ time series, respectively. We continued moving the SW one unit to the right at each iteration and calculated the corresponding
correlation until the final iteration (that is, when the right-end of the SW is at $T = 300^{th}$ time point).  The performance of this method depends on the size of the window: a larger $ws$ provides smoother estimates of dynamic correlation, while smaller $ws$ is more sensitive to variations in the underlying data. Lindquist \textit{et al} [8] discusses this issue in more detail, and also mentions variations of the above technique such as tapered sliding window [1].

\subsection*{Method $\#$ 2 - Dynamic conditional correlation}

DCC is a multivariate volatility model commonly used in financial analyses, which also has neuroimaging applications. Lindquist \textit{et al} provides an accessible exposition of DCC using mathematical notation. The original papers by Engle and Sheppard [9, 10] are
also good sources on DCC. As mentioned in Lindquist \textit{et al}, ``DCC provides a parametric approach towards estimating dynamic correlations,
much like auto-regressive (AR) and auto-regressive moving average (ARMA) models provide a parametric approach towards modeling fMRI noise [12]". In
particular, DCC is based on generalized autoregressive conditional heteroscedastic (GARCH) models [13], which were developed to incorporate the so-called ``stylized facts" (i.e. statistical irregularities) common to a large number of financial time series [14]. The key feature of DCC that we would like to emphasize here is that it is parametric. The parameters in DCC are estimated using a two step ML estimation approach, the mathematical details of which can be found in Lindquist \textit{et al}. We will not repeat this description but instead we will explain how DCC estimation can be performed using the R packages \textit{ccgarch} and \textit{rmgarch}.

Two R packages - \textit{ccgarch} and \textit{rmgarch} - have been developed for applying the DCC method. The main function for DCC estimation in
\textit{ccgarch} is called `\textit{dcc.estimation}'. The inputs required for this R function are the initial values for the GARCH parameters (see
equations 19 and 20 in Lindquist \textit{et al}) [8]; the specification of the numerical optimization routine for maximizing the log-likelihood; and
the two time series whose dynamic correlation is to be estimated. In our own analyses, we used the Broyden-Fletcher-Goldfarb-Shannon (BFGS) algorithm and the diagonal model for GARCH covariance matrix mentioned in Lindquist \textit{et al.}, specified using \textit{model = ``diagonal"} argument within the \textit{dcc.estimation} function. Among the many outputs provided
by \textit{dcc.estimation}, the most relevant for the current work is the one named \textit{DCC} which stores the dynamic conditional correlation as
a time series vector in R.

The main function for DCC estimation in \textit{rmgarch} is called `\textit{dccfit}'. This function requires three inputs: specification of the
model, the multivariate time series, and a solver to be used for optimization. In our work, we specified the model by using a separate command called \textit{`dccspec'}; the GARCH part of the specification was accomplished by using \textit{ugarchspec}. The optimization method employed in all our simulations and examples was the default named \textit{`solnp'} which performs nonlinear optimization using augmented Lagrange multiplier method
[15].

A key point is that in the original DCC model presented by Engle and Sheppard [9, 10] multivariate normality
of the underlying pair of time series was assumed for ML estimation. As mentioned in the same papers [9, 10], multivariate normality is not necessary for statistical consistency of the estimators, as the estimation can be extended to a quasi-likelihood framework. However, even in the
quasi-likelihood framework, certain assumptions (e.g. related to the moments) are often necessary. Such assumptions may not be valid for time series
which contain extreme values.


\subsection*{Method $\#$ 3 - Nonparametric weighted VGA (WVGA)}

As mentioned above, DCC depends on assumptions of normality or at least finiteness of moments. To avoid dependence of such assumptions, in our new method both time series were converted into graphs using the weighted visibility graph algorithm (WVGA), introduced in Supriya \textit{et al} [16]. WVGA, developed by Lacasa \textit{et al} [11], is an extension of the visibility graph algorithm that converts simply but elegantly a time series to a graph. Methods based on visibility graphs have become popular after the original concept of mapping time series to a complex network was introduced in Zhang and Small [17]. The nodes of the WVGA-based graphs correspond to the time points and therefore, the graph generated from each time series consists of $T$ nodes. The weight $w_{ab}$ of the edge between the nodes corresponding to time points $t_{a}$ and $t_{b}$, with $a < b$, is calculated using equation (2) given in p.6558 in Supriya \textit{et al} [16]:
\[ w_{ab} = \arctan{ \frac{x(t_{b}) - x(t_{a})}{t_{b} - t_{a}}}. \] Here $x(t_{a})$ and $x(t_{b})$
are the values of the time series at time points $t_{a}$ and $t_{b}$,
respectively. All edge weights are calculated in radians.

We modified WVGA in several ways. First, strictly speaking, for the graph to be a visibility graph, there should be no edge between two nodes if
there is no ``visibility" between them, where ``visibility" is defined based on the criterion spelled out in Lacasa \textit{et al.}[11]. However, in our
version of the algorithm we kept all the edges, and each edge had an assigned weight based on the arctan function mentioned above. Since this
algorithm is very similar to the one proposed in Supriya \textit{et al.} [16], we retain the name "weighted visibility graph algorithm". Second, while
Supriya \textit{et al.} [16] used only the absolute values of the weights, we retained the signs of the weights calculated, as is. The reason we
introduced these modifications is that the shape of the arctan function, rather than the visibility criterion, provides a suitable solution for our
purpose. We elaborate more on this in the discussion section.

Our modified WVGA for dynamic correlation estimation is also based on a sliding window. As for method 1, we picked a window of size 15 for
illustration. Associated with each time point $t_{i}$ (that is, with each node $n_{i}$ in the WVGA graph) is a weight-vector of length $T$ consisting
of the weights of the edges from all the other nodes of the graph to $n_{i}$. We named this weight vector as $\mathbf{w}_{i}$. Note that we
considered the weight between $n_{i}$ to itself as zero. That is, $i^{th}$ element of $\mathbf{w}_{i}$ is zero, for all $i$. For each time series, we
place the right-end of a sliding window (of size 15) at time point $i$, considered all the weight vectors \[ \left\{\mathbf{w}_{i-15 + 1},
\mathbf{w}_{i-15+2}, \ldots, \mathbf{w}_{i} \right\}, \] and obtained a new vector $\mathbf{W}_{(i:\,\mathrm{median})}$ by taking the median
element-wise. That is $k^{th}$ element of $\mathbf{W}_{(i:\,\mathrm{median})}$ is the median of the $k^{th}$ elements of $\left\{\mathbf{w}_{i-15 +
1}, \mathbf{w}_{i-15+2}, \ldots, \mathbf{w}_{i} \right\}.$ If we use super-scripts to denote the vectors corresponding to the two time series, then
the above steps may be summarized using the following formula: \[\mathbf{W}^{(j)}_{(i:\,\mathrm{median})}[k] =
\mathrm{median}{\left\{\mathbf{w}^{(j)}_{i-15+1}[k], \mathbf{w}^{(j)}_{i-15+2}[k], \ldots, \mathbf{w}^{(j)}_{i}[k] \right\}, } \]  where $k ( = 1,
\ldots, T)$ denotes the $k^{th}$ element of each vector and $j ( = 1, 2)$ denote the $j^{th}$ time series. The correlation at the $i^{th}$ iteration
(where $i$ ranges from 15 to $T$) is then just the correlation between the vectors $\mathbf{W}^{(1)}_{(i:\,\mathrm{median})}$ and
$\mathbf{W}^{(2)}_{(i:\, \mathrm{median})}$.

To illustrate graphically how the three methods perform, we computed the three dynamical correlations and plotted them together in the bottom panel
of Figure 1. The underlying dynamic correlation, which is equal to zero at all time points, is plotted as the black line. The correlations obtained through the SW method are plotted as the blue line. There are two
superimposed green lines corresponding to the two R packages used for DCC estimation - \textit{ccgarch} and \textit{rmgarch}. The lines superimpose
because in this case the two results coincide. However, this is not always the case with real data example, as it will be seen below. The green curves may appear to be straight lines, but this is not in fact the case (see Figure A1 in the Appendix
where we used a different scale for the y-axis). The correlations obtained through the nonparametric WVGA method are plotted as the magenta curve.

SW method (blue curve) results in the worst estimate, DCC method (green curves) results in the best estimate, and the nonparametric WVGA method
(magenta curve) performs somewhere in between. This qualitative estimation is confirmed quantitatively by the following summary statistics: mean
absolute value of the blue curve is 0.19, for both green curves is 0.06, and for the magenta curve is 0.10; the maximum value for the blue, (both) green, and magenta curves are, respectively, 0.61, 0.06, and 0.39. In the above example, DCC performs better than both the SW technique and our new
method. However, the data in this example are a pair of nicely behaved time series with no outliers (since they were generated from a bivariate
normal density). In the presence of outliers DCC does not perform as well. We generated two time series from a bivariate Cauchy density (by
using the R function \textit{rcauchy2d} from the package \textit{fMultivar}). Because Cauchy density has very heavy tails, the generated data contain several extreme values.The correlation parameter for the bivariate Cauchy density ($\rho$) was
set to zero, so that the underlying dynamic correlation for this example was also zero for all time points. The two time series thus generated are
plotted in the top panel in Figure 2. We applied the
same three analytical methods to this data set and the results are plotted in the bottom panel in Figure 2, with the same color coding as in Figure
1. For this example, the new method easily outperformed the other two methods. The mean absolute values for the blue, green, and magenta curves were,
respectively, 0.50, 0.46 (for both green curves) and 0.35; the corresponding maximums of the absolute values of the correlations were 0.98, 0.96 (for
both green curves) and 0.67, respectively. Thus, the new method, nonparametric in nature, was more robust when extreme values were present in the data. As a minor note, the y-axes for the top panels had the same scale for Figures 1 and 2, which cut off the plot for some extreme values in Figure 2. In order to see the actual extreme values, we plotted the same pair time series with extended y-margins in Appendix Figure A2 .

\begin{figure}[H]
\begin{center}
\includegraphics[height=3in,width=7in,angle=0]{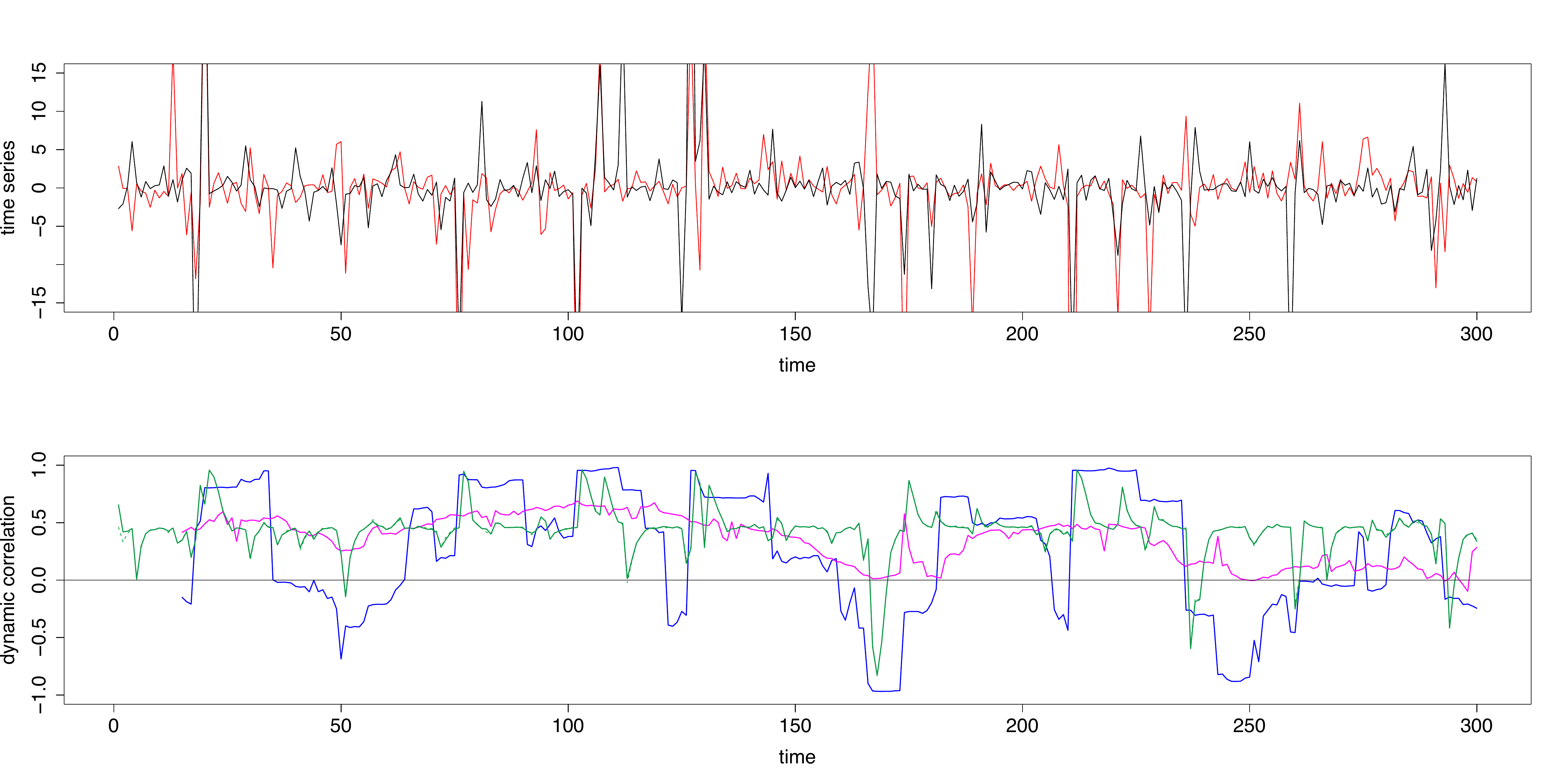}
\caption{Top panel: The two time series used for the second illustrative simulated example; data
were generated
from a bivariate Cauchy distribution. Bottom Panel: Dynamic correlations estimated for the
above two time series using various methods. Method $\#$1, SW (blue), method $\#$2, DCC (green),
and method $\#$3, based on modified WVGA (magenta). The black horizontal line in the center
through zero represents the underlying true dynamic correlation for this example.}
\end{center}
\end{figure}

\subsection*{Simulation design}

To assess the performance of the three methods under various scenarios, we ran extensive simulations using scenarios aimed to closely match the work
of Lindquist \textit{et al}[8]. We therefore generated data by using bivariate normal distribution, as in Lindquist \textit{et al}[8], but also
bivariate Cauchy distribution, to obtain data sets with extreme values. In all the simulations involving bivariate Cauchy, values exceeding 50 were
reset to 50, and values below -50 were reset to -50. The means for all pairs of time series in the simulation study was set to $(0,0)$ across all
time points and the covariance matrix at time point $t$ was
                           \[  \left[ \begin{array}{cc} 2     & \sqrt{6}p(t) \\
                                                 \sqrt{6}p(t) & 3 \end{array} \right].  \]
By considering three different forms for $p(t)$ we obtained three different simulations scenarios,
as in Lindquist \textit{et al} [8]. For simulations involving the sliding window technique (method1)
and the nonparametric WVGA (method3), the window size was set to 15. \\

\textit{Simulations Design D1:} In this case, we set $p(t) = 0$ for all $t$ as in the illustrative examples above. We considered four time series of
different lengths ($T = 150, 300, 600$ and $1000$) for both bivariate normal and bivariate Cauchy distributions. The metrics used to assess the
performance of the methods for this
scenario were the mean and maximum across all time points. \\

\textit{Simulations Design D2:} The scenarios in $D2$ were
designed to assess the performance of the methods when the underlying correlation varies
slowly over time. We set $p(t) = \sin\left(t/\Delta \right)$, $\Delta = 1024/2^{k}$,
with $k = 3, 4$. We obtained two different (sub-)scenarios corresponding to $k = 3$ and $4$, which we
refer to as $D2a$ and $D2b$. We set $T = 600$ for both of these sub-scenarios.  The metric
used to assess the performance was mean-squared-error (MSE) as in Lindquist \textit{et al}.  \\

\textit{Simulations Design D3:} The scenarios in $D3$ were
designed to assess the performance of the methods when there is a rapid change in the underlying
correlation; here, the rapid change occurs near time point 250.
We set $p(t)$ equal to a Gaussian kernel with mean 250
and standard deviation 15$k$, where $k$ = 3 and 4. The two $k$'s resulted in two different (sub-)scenarios,
$D3a$ and $D3b$;  $T = 600$ for both cases.  MSE was used to compare the performance of the methods.

\subsection*{Real data analysis}

To illustrate the advantages and disadvantages of the methods considered in this paper, we applied our analysis to two different empirical data sets.
The first data set used for illustration was fMRI data collected from one healthy adult subject while the subject performed a color-word Stroop task
[18]. The data were part of a sample collected for a study of the underlying mechanisms based on distributed network of brain regions involved in
congruency sequencing effect in cue-conflict paradigms [19]. Time series from six different brain regions were used in our analysis, and dynamic
correlations between each pair of regions were obtained. The second data set used for illustration consisted of local field potential (LFP) time
series [20, 21] obtained from four different electrodes implanted in the brain of a rat. The recording occurred while the rat was awake and placed on
a small platform in the lab. Three of the recordings were from the CA1 field of the hippocampus, and one recording from the medial dorsal striatum.
Dynamic correlations of data from each pair of recordings were computed. Although the behavior of the dynamic correlations of pairwise time series
was of intrinsic interest for the background scientific research question related to fMRI and LFP data sets, the main purpose in the current
analysis was to illustrate various aspects associated with the estimation methods considered in this paper. Specifically, the Stroop task fMRI
data set illustrated the convergence issues related to DCC and LFP data illustrated the performance of the methods in the presence of extreme values.
Details about data extraction for fMRI data set and recording for the LFP data are presented below.

\textit{fMRI data}. Sample fMRI data from one healthy human adult subject were downloaded from OpenfMRI.org. We used Stroop task fMRI data
(https://openfmri.org/dataset/ds000164/) [19], since activation patterns for the Stroop task are relatively well known. Preliminary data processing
followed previous publication [22]. Using FMRIB Software Library (FSL) as well as Analysis of Functional NeuroImages (AFNI), anatomical volume was
skull stripped, segmented (gray matter, white matter and CSF), and registered to the MNI 2mm standard brain. Processing of fMRI echo planar image
(EPI) volume included the following. The first four EPI volumes were removed; transient signal spikes were removed; head motion was corrected; the
volumes were smoothed with a 6mm FWHM Gaussian kernel; the volumes were resampled, spatially transformed and aligned to the MNI 2mm standard brain
space; volumes with excess motion were scrubbed. In order to extract the time series for six regions chosen {\it a priori}, the frontal medial cortex,
subcallosal frontal region, insular cortex, Heschl's gyrus, and amygdala were defined using the Harvard-Oxford atlas and the Caudate Head was defined
using WFU PickAtlas.The EPI time series was extracted within each of these six regions.

\textit{LFP data}. The local field potential (LFP) data were recorded from an awake rat placed on a small platform. Previous to the recording, the
animal had been implanted with a 16 tetrode hyperdrive assembly intended for recording in the CA1 field of the hippocampus, combined with two
cannula/electrode systems (PlasticsOne, Inc) aimed bilaterally at the medial dorsal striatum. The tetrodes were gradually lowered in the CA1 layer
across 8 to 10 days after recovery from surgery, and their tip was positioned based on the configurations of sharp waves and ripples. The electrodes
aimed at the dorsal striatum were implanted directly in the position from which the recording was obtained (AP: + 2.5mm, ML: +/-2.4mm; DV: -5.4mm;
tilted 22 degrees from the vertical in the sagittal plane). The single wire electrodes located in the striatum and the 16 tetrodes were connected
through a PCB to a group of four headstages with a total of 64 unity gain channels and 2 color LEDs for position tracking. The LFP signals were
amplified (1000x), band-pass filtered (1-1000Hz), digitized (30KHz), and stored together with LED positions on hard disk (Cheetah Data Acquisition
System, Neuralynx, Inc.).  Subsequently the signal was downsampled to 1KHz, in which form it was used for the current analysis.

\section*{Results}

\subsection*{Simulations}

\textit{Simulations Design D1:} Results from simulations with design D1 are presented in tables 1 and 2. Table 1  presents the results when the pair
of time series was generated from a bivariate normal distribution and table 2 presents the results related to bivariate Cauchy. Note that one of the two DCC estimates consistently had convergence problems. In the bivariate
normal case, across all T values, DCC performed  best, SW performed worst, and nonparametric WVGA's performance was in between. In the bivariate
normal case, DCC’s performance improved with larger T as measured both by bias and variance, while that of SW and nonparametric WVGA methods remained
the same for all T. In the bivariate Cauchy case, the SW  method performed consistently worst across all T, while for the other two methods
performance  improved with larger T. At small T values, the nonparametric WVGA method performed better than the DCC method. As T became larger, the DCC
performance overtook that of WVGA, and at T = 1000,  DCC performed better than WVGA even if compared by using MSE: [$(0.200^2) + (0.035^2)] \;=\;
0.041$ versus [$(0.126^2) + (0.084^2)] \;=\;  0.023$. Thus, in the presence of extreme values, DCC was robust  as long as the number of time points was
large enough; when the number of time points was small, WVGA-based evaluation is preferable to DCC.

\begin{center}
  \tabcolsep=0.11cm
    \begin{tabular}{ | l || c | c | c | c | }                             \hline
        \multicolumn{5}{|c|}{Table 1. Bivariate Normal in design D1}   \\ \hline
        \multicolumn{5}{|c|}{Mean of absolute value of correlations across time}   \\ \hline
                         & T = 150            & T = 300            & T = 600  & T = 1000    \\ \hline
         SW              &  $0.218 (0.039)$   & $0.217 (0.027)$    & $0.218 (0.019)$  & $0.218 (0.015)$    \\ \hline
         WVGA-based      &  $0.134 (0.033)$   & $0.129 (0.021)$    & $0.127 (0.015)$  & $0.125 (0.012)$    \\ \hline
         DCC1            &  $0.071 (0.053)$   & $0.051 (0.037)$    & $0.037 (0.025)$  & $0.029 (0.020)$    \\ \hline
         DCC2            &  $0.077 (0.052)$   & $0.055 (0.037)$    & $0.040 (0.025)$  & $dnc^{*}$    \\ \hline
         \multicolumn{5}{|c|}{Maximum of absolute value of correlations across time}   \\ \hline
                         & T = 150            & T = 300            & T = 600  & T = 1000    \\ \hline
         SW              &  $0.616 (0.094)$   & $0.670 (0.075)$    & $0.717 (0.065)$  & $0.744 (0.053)$    \\ \hline
         WVGA-based      &  $0.392 (0.080)$   & $0.425 (0.067)$    & $0.455 (0.058)$  & $0.475 (0.054)$    \\ \hline
         DCC1            &  $0.093 (0.101)$   & $0.075 (0.087)$    & $0.067 (0.077)$  & $0.055 (0.068)$    \\ \hline
         DCC2            &  $0.139 (0.140)$   & $0.111 (0.115)$    & $0.104 (0.105)$  & $dnc^{*}$    \\ \hline \hline
         \multicolumn{5}{|c|}{$*dnc$ = did not converge}   \\ \hline
    \end{tabular}
\end{center}

\begin{center}
  \tabcolsep=0.11cm
    \begin{tabular}{ | l || c | c | c | c | }                             \hline
        \multicolumn{5}{|c|}{Table 2. Bivariate Cauchy in design D1}   \\ \hline
        \multicolumn{5}{|c|}{Mean of absolute value of correlations across time}   \\ \hline
                         & T = 150            & T = 300            & T = 600  & T = 1000    \\ \hline
         SW              &  $0.531 (0.080)$   & $0.529 (0.057)$    & $0.526 (0.040)$  & $0.528 (0.030)$    \\ \hline
         WVGA-based      &  $0.244 (0.098)$   & $0.222 (0.067)$    & $0.210 (0.048)$  & $0.200 (0.035)$    \\ \hline
         DCC1            &  $0.292 (0.188)$   & $0.224 (0.143)$    & $0.168 (0.111)$  & $0.126 (0.084)$    \\ \hline
         DCC2            &  $dnc*$            & $dnc$              & $dnc$            & $dnc$    \\ \hline
         \multicolumn{5}{|c|}{Maximum of absolute value of correlations across time}   \\ \hline
                         & T = 150            & T = 300            & T = 600  & T = 1000    \\ \hline
         SW              &  $0.973 (0.030)$   & $0.986 (0.012)$    & $0.992 (0.005)$  & $0.994 (0.003)$    \\ \hline
         WVGA-based      &  $0.540 (0.121)$   & $0.558 (0.097)$    & $0.576 (0.081)$  & $0.590 (0.074)$    \\ \hline
         DCC1            &  $0.363 (0.279)$   & $0.321 (0.281)$    & $0.302 (0.289)$  & $0.245 (0.267)$    \\ \hline
         DCC2            &  $dnc$             & $dnc$              & $dnc$            & $dnc$    \\ \hline \hline
          \multicolumn{5}{|c|}{$*dnc$ = did not converge}   \\ \hline

    \end{tabular}
\end{center}

\begin{figure}[H]
\begin{center}
\subfloat{\includegraphics[height=2in, width = 3in]{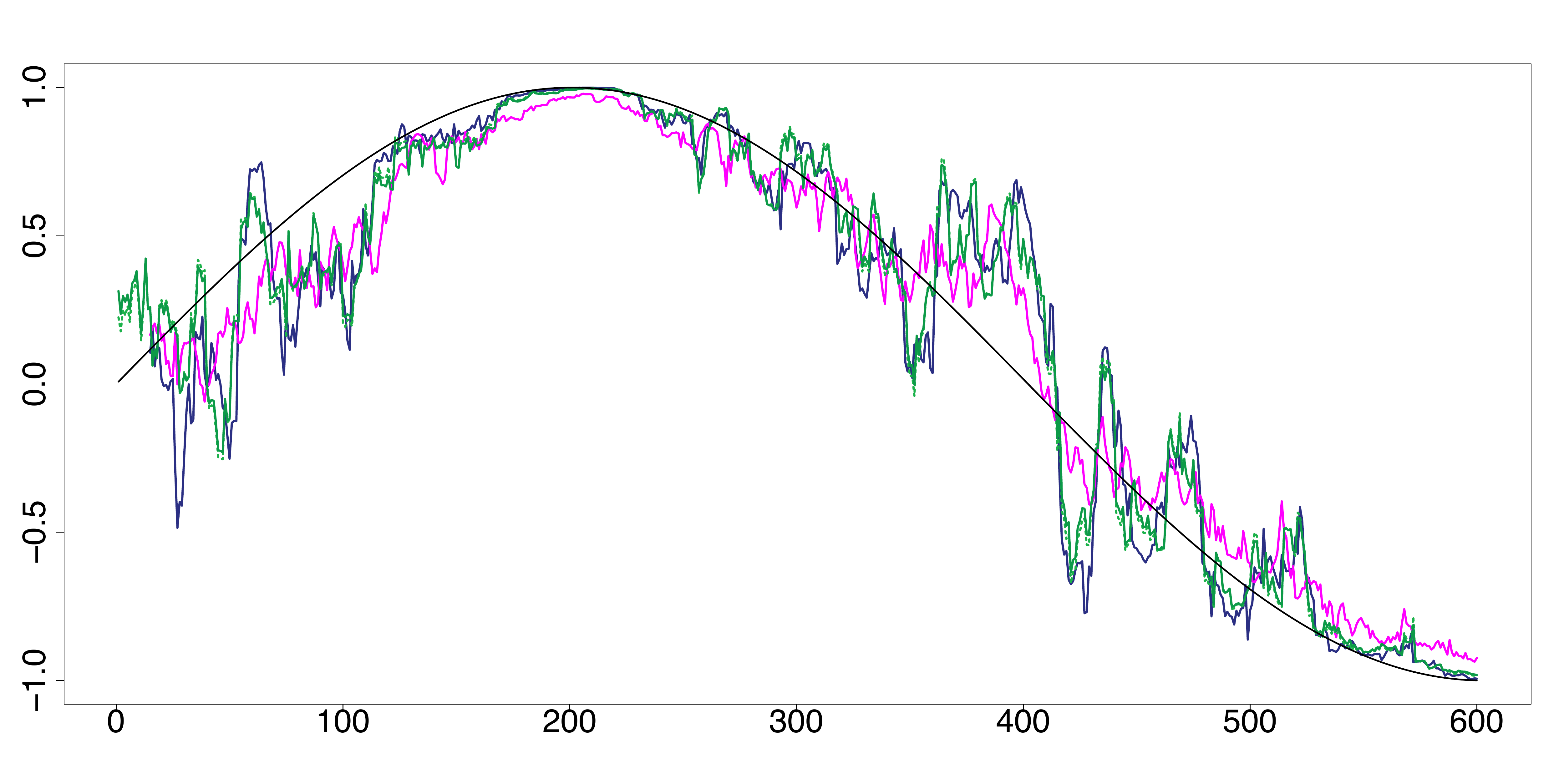}}
\subfloat{\includegraphics[height=2in, width = 3in]{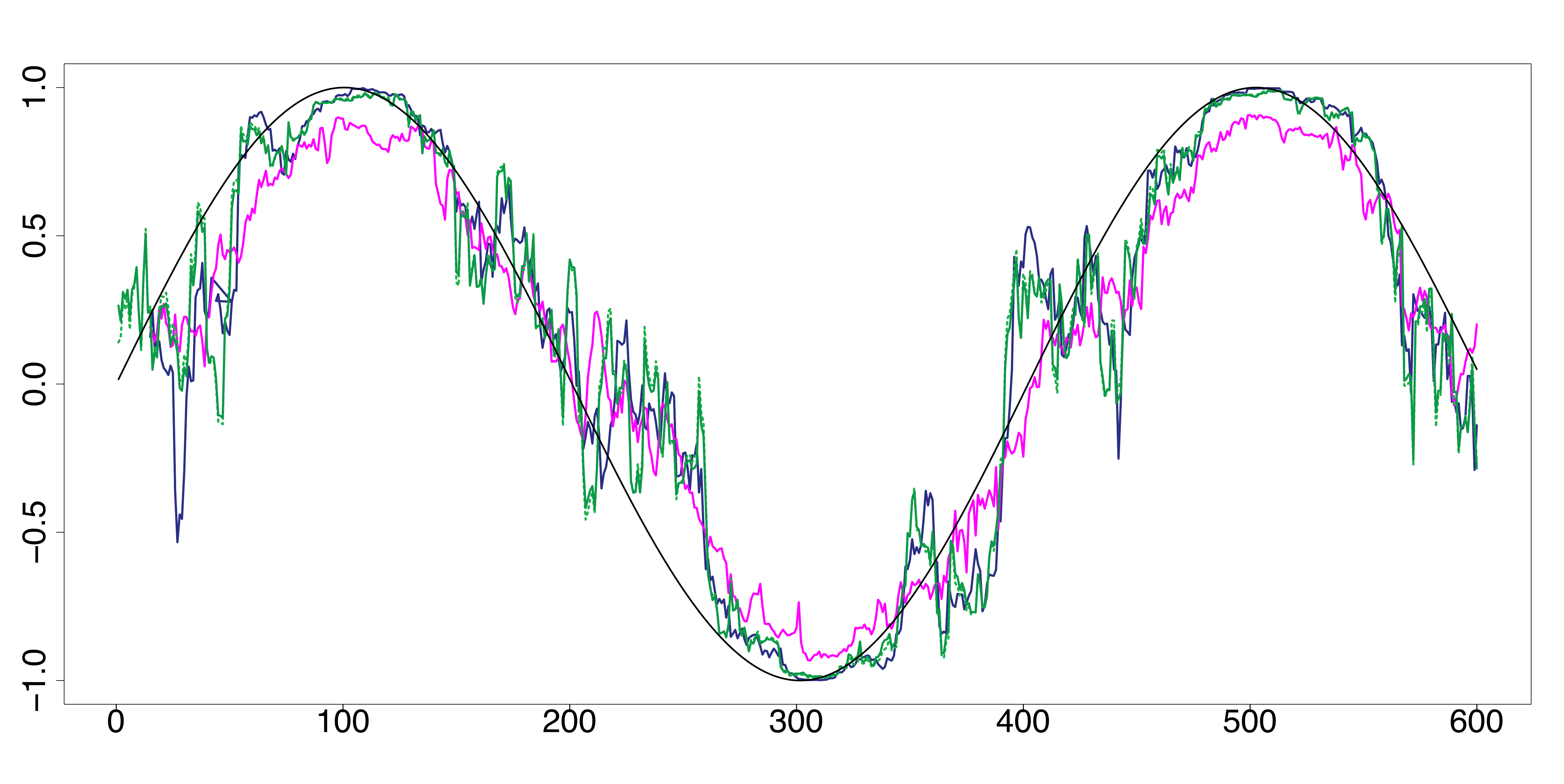}} \\
\subfloat{\includegraphics[height=2in, width = 3in]{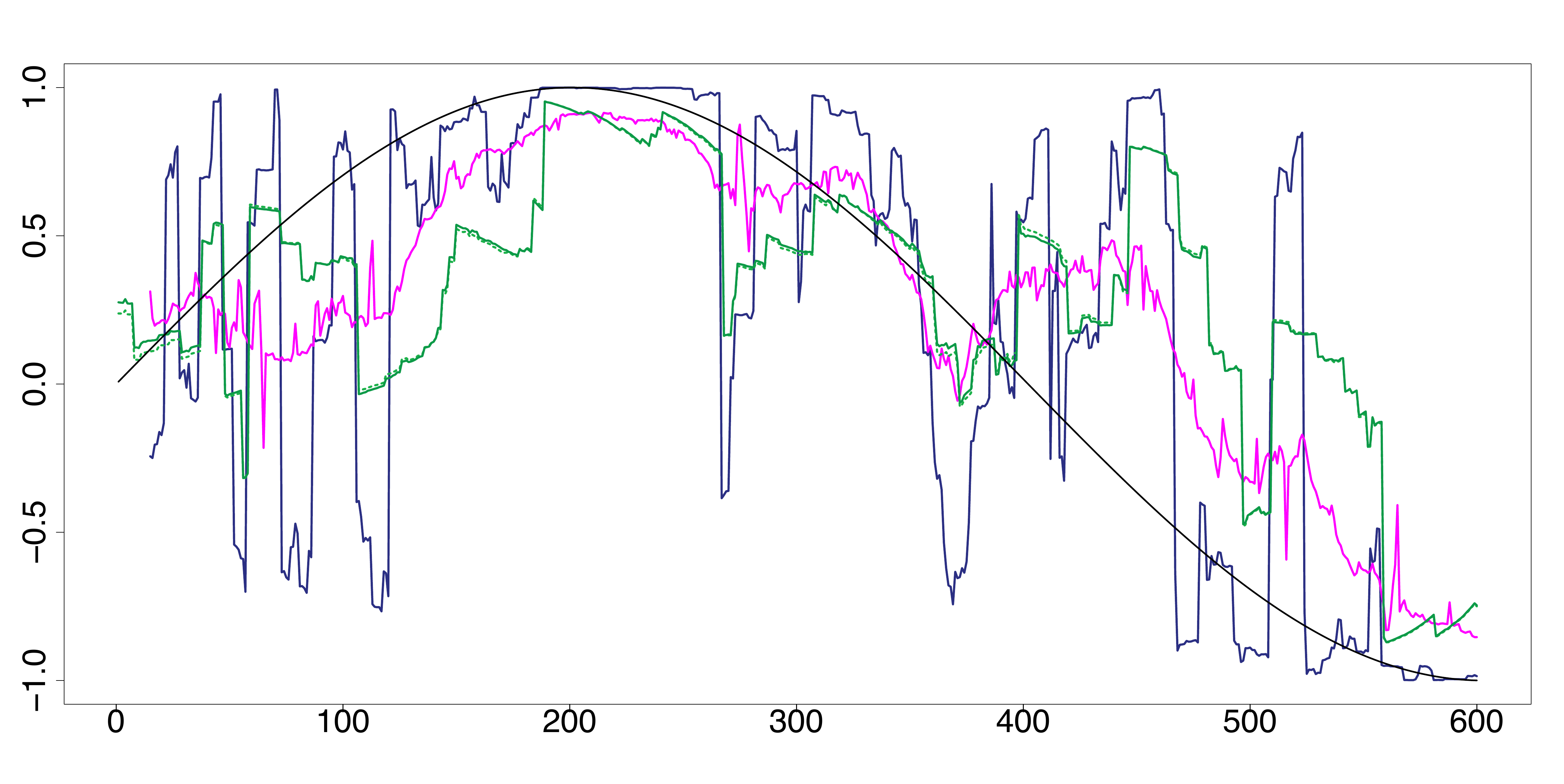}}
\subfloat{\includegraphics[height=2in, width = 3in]{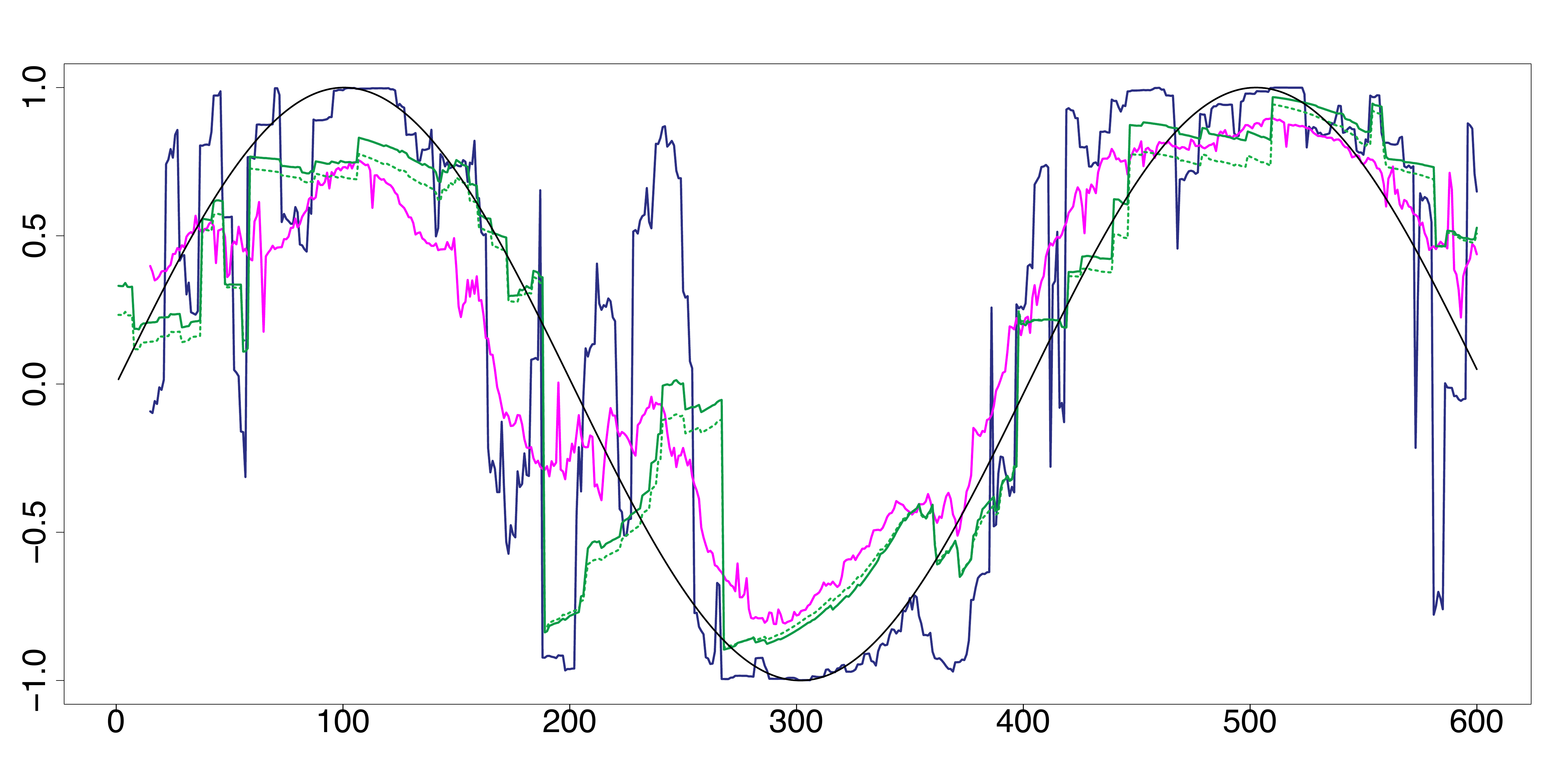}} \\
\caption{Results for single iteration from simulations design D2.
Top row: Underlying pair of time series from bivariate normal.
Bottom row: Underlying pair of time series from bivariate Cauchy.
Left column: Design 2a ($k=3$). Right column: Design 2b ($k=4$).
The underlying true dynamic correlations are plotted as the black curve.
Blue (SW), magenta (WVGA), green solid (DCC with \textit{rmgarch})
and green dashed (DCC with \textit{ccgarch}) represents the estimates
of dynamic correlation. }
\end{center}
\end{figure}

\begin{figure}[H]
\begin{center}
\subfloat[]{\includegraphics[height=2in, width = 2in]{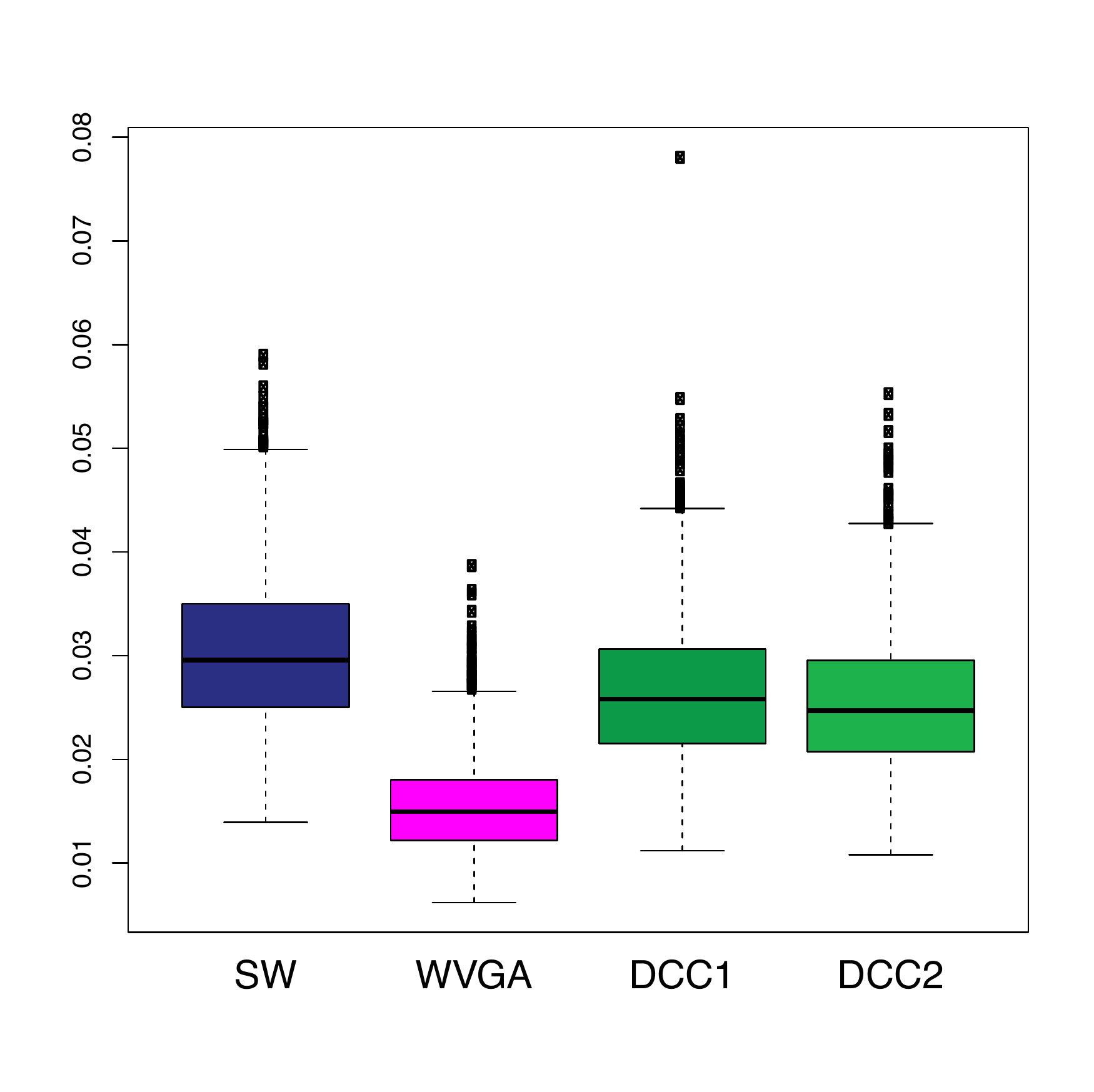}}
\subfloat[]{\includegraphics[height=2in, width = 2in]{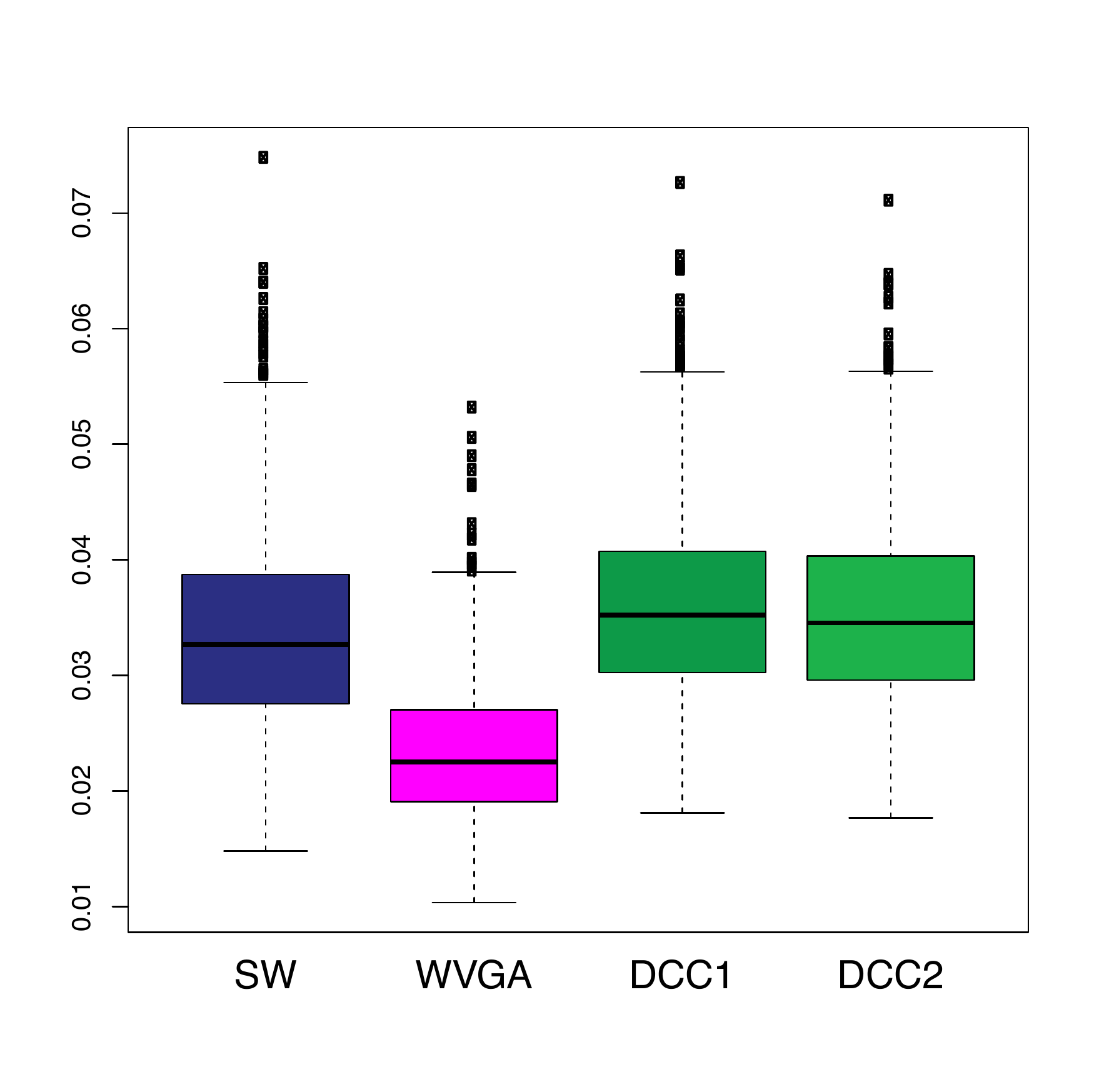}}\\
\subfloat[]{\includegraphics[height=2in, width = 2in]{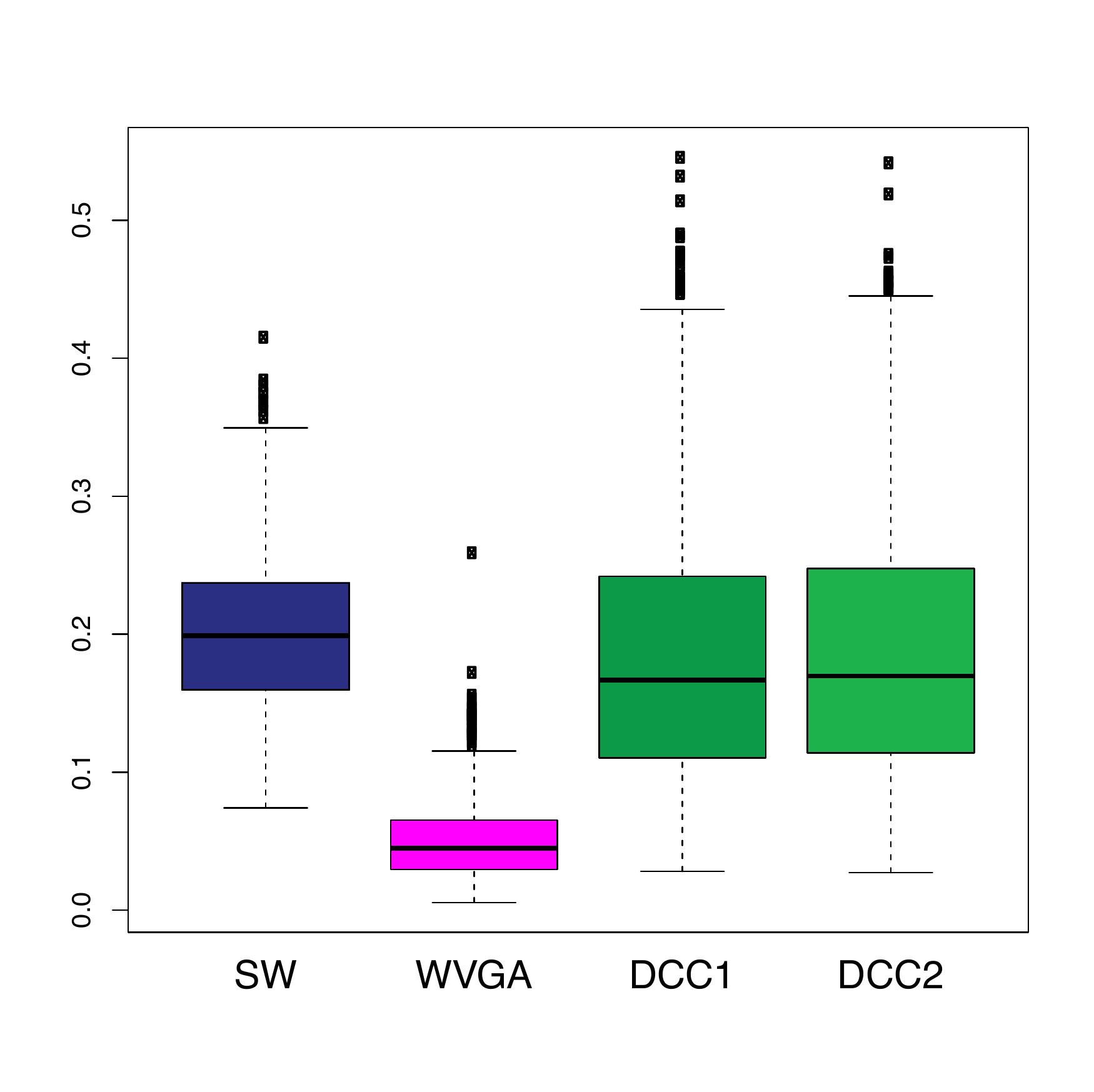}}
\subfloat[]{\includegraphics[height=2in, width = 2in]{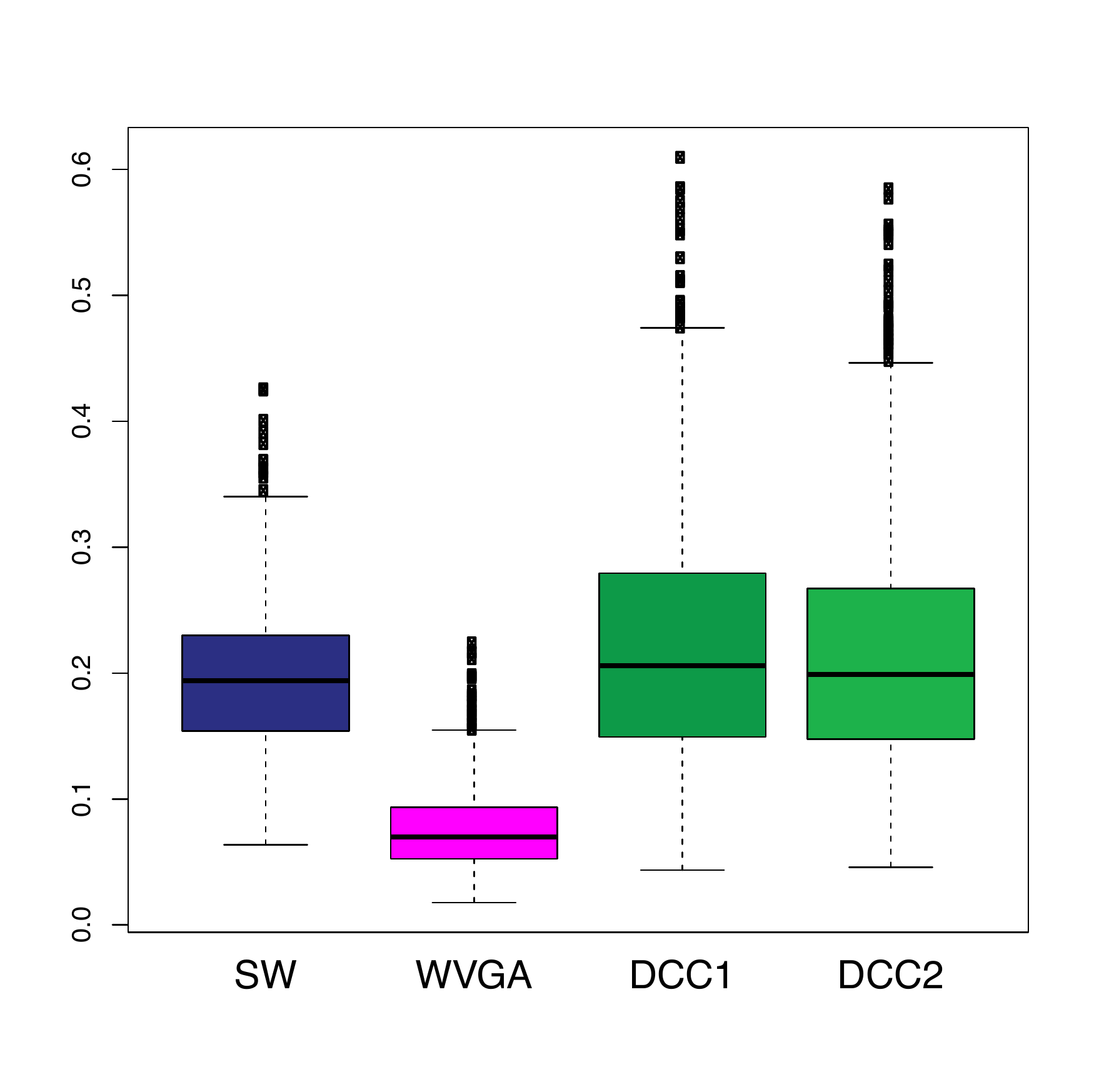}}
\caption{Boxplots of mean squared error of the estimation methods based
on 1000 iterations from simulations design D2. The panels correspond to
bivariate Normal (top row), bivariate Cauchy (bottom row), design D2a (left
column) and design D2b (right column), as in figure 3.}
\end{center}
\end{figure}

\textit{Simulations Design D2:} In Figure 3, the black curves in the left panels are the underlying true dynamic correlations corresponding to
simulations design D2a, the black curves in the right panels correspond to that of the simulations design D2b. Each panel refers to a single simulation iteration. The estimates of the dynamic
correlations obtained via the three methods presented in this paper are plotted as colored curves. The top panels
correspond to the bivariate normal scenario and the bottom panels correspond to the bivariate Cauchy scenario. Results from 1000 simulations for each
scenario are plotted in Figure 4. In this design, the performance of the WVGA-based method was far superior to the other two methods for both bivariate normal and
bivariate Cauchy simulations.

\textit{Simulations Design D3:} Figure 5 plots the underlying true dynamic correlations and the estimates based on the three methods for a single
iteration corresponding to the simulation design D3. Interestingly, the DCC estimates based on both R packages (the green straight lines) did not
capture the dynamic shift in correlations at all. The results based on 1000 iterations are plotted in Figure 6. In all four scenarios for the design
(D3a and D3b, bivariate normal and Cauchy), the performance of SW was substantially worse compared to the other two methods. DCC's performance was
comparable to that of nonparametric WVGA method in the bivariate normal scenario, although the WVGA method had a slight edge over DCC. In the
bivariate Cauchy case (bottom panels) the improvement in performance for the WVGA-based method is more substantial. Thus, the results of the
simulations indicate that when the underlying dynamic correlations are not constant across time, nonparametric WVGA performed better than DCC in many
of the scenarios, and never worse than DCC in any of the scenarios considered.

\begin{figure}[H]
\begin{center}
\subfloat{\includegraphics[height=2in, width = 3in]{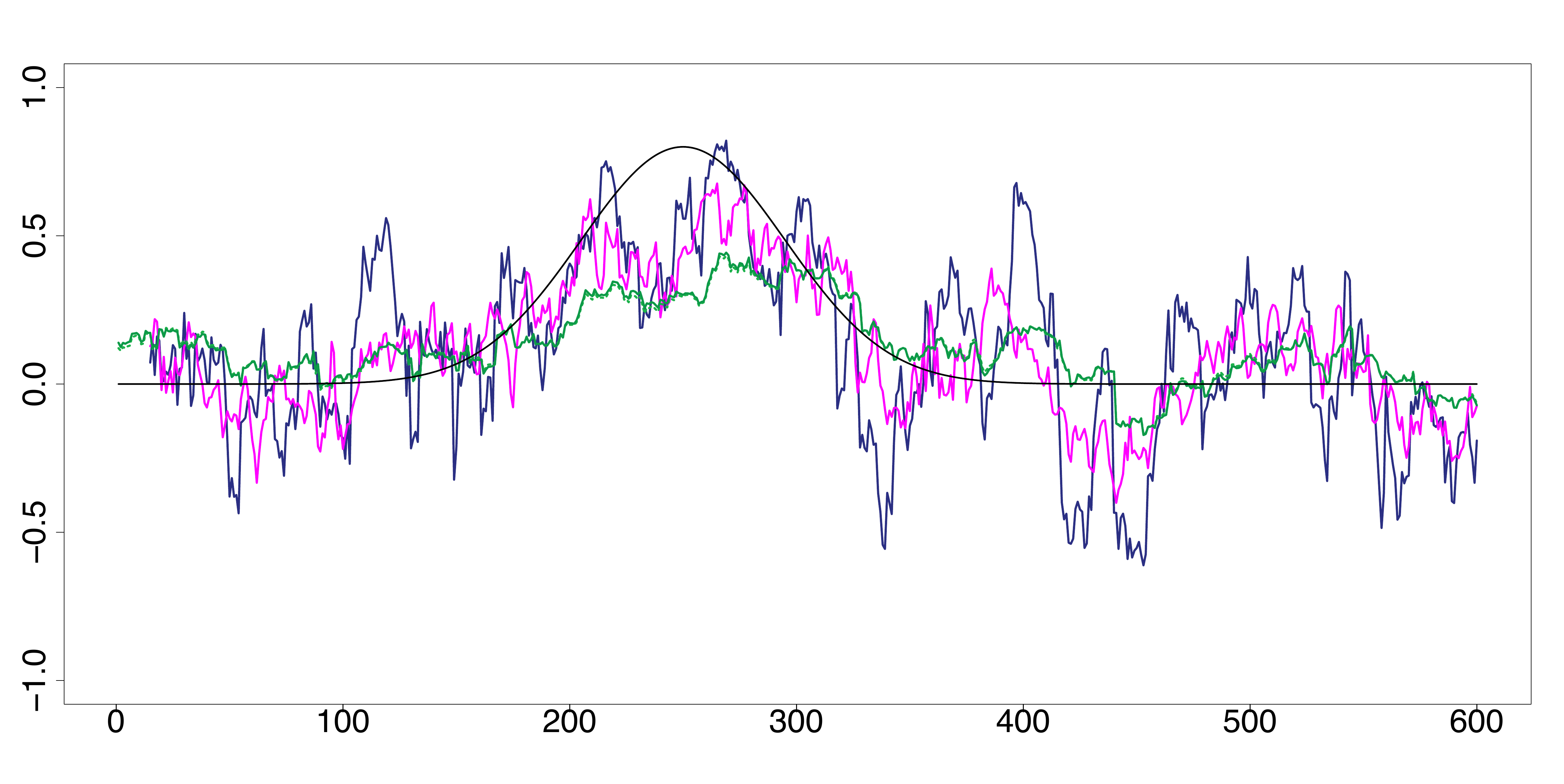}}
\subfloat{\includegraphics[height=2in, width = 3in]{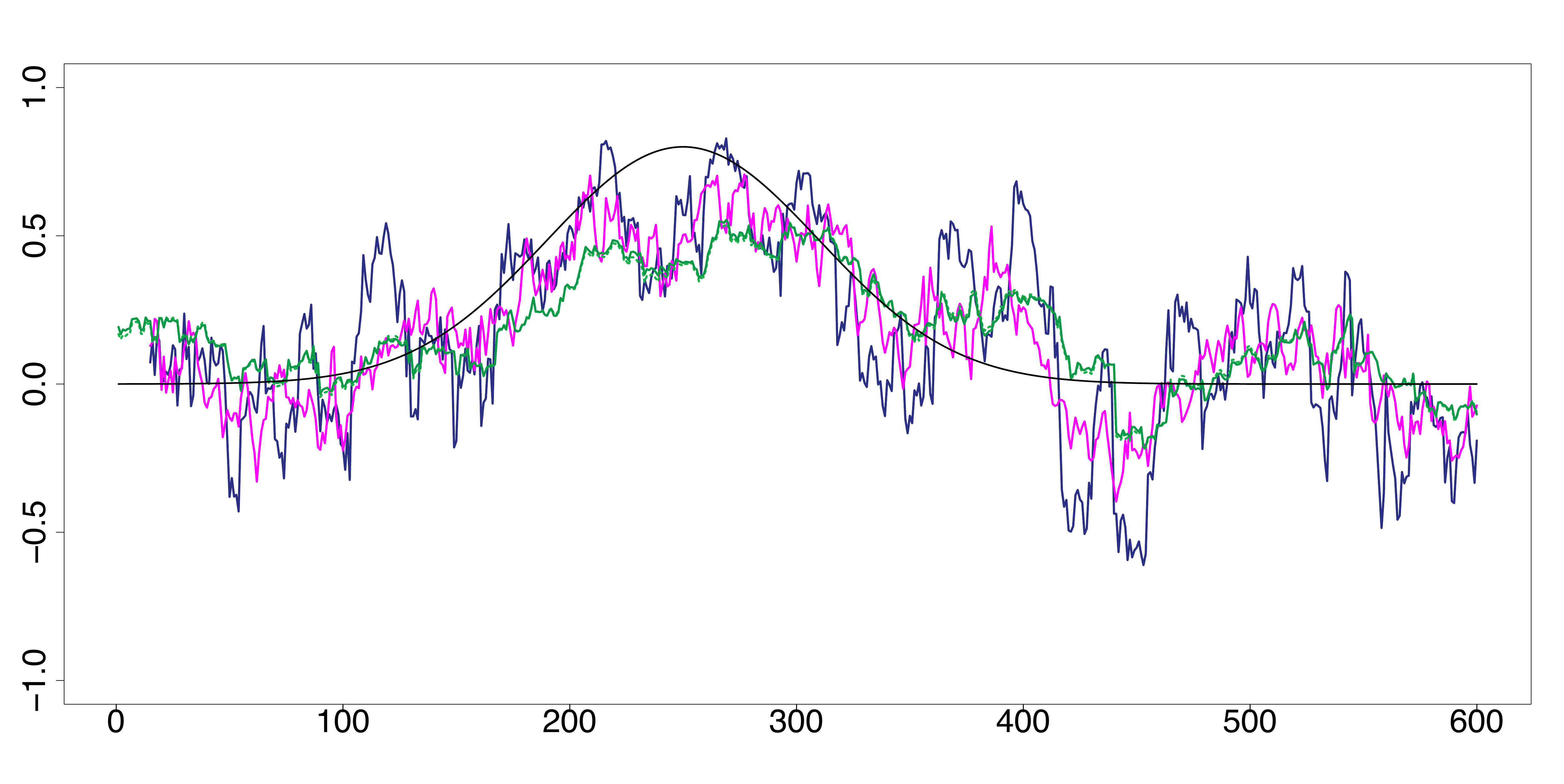}} \\
\subfloat{\includegraphics[height=2in, width = 3in]{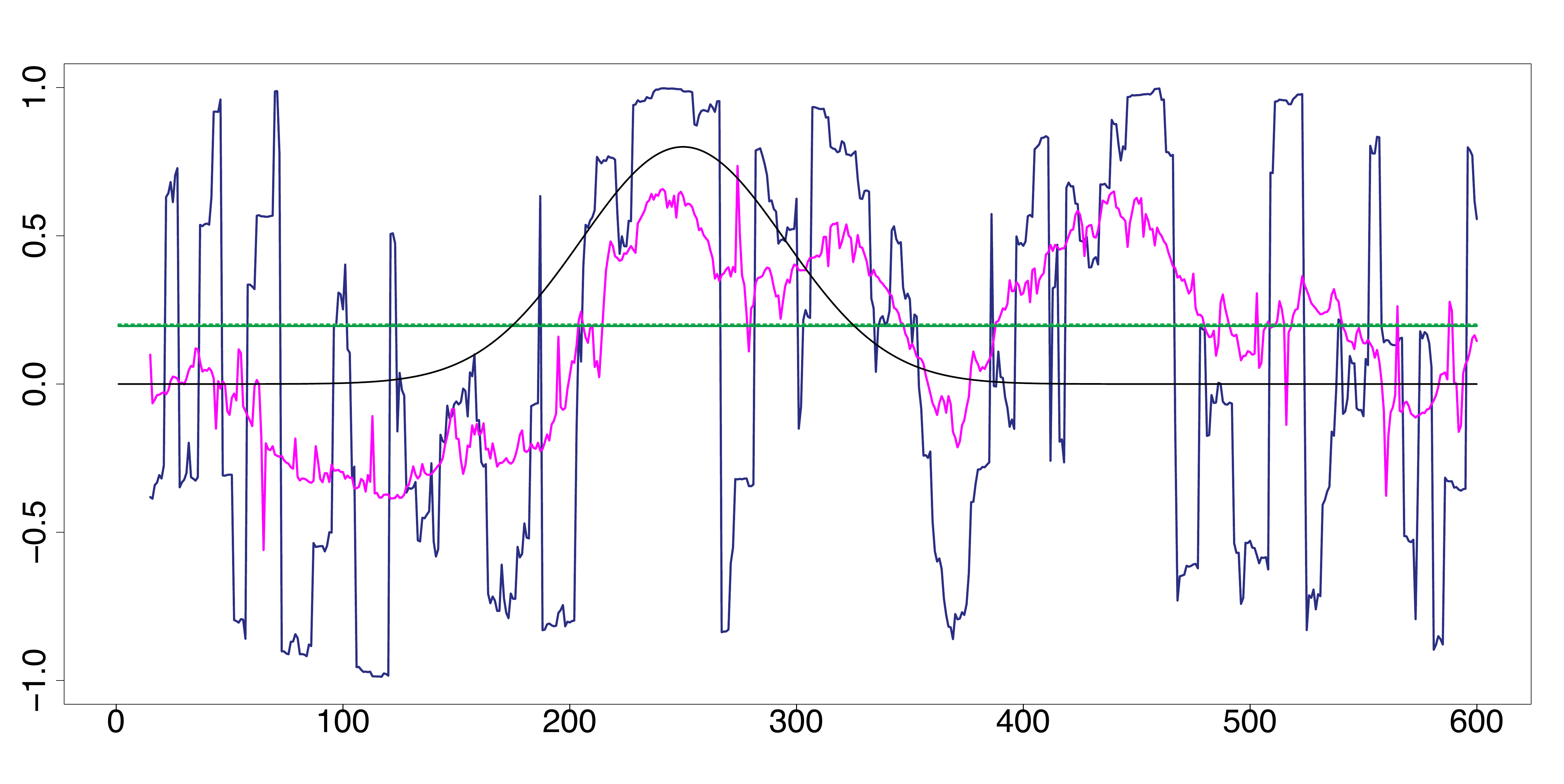}}
\subfloat{\includegraphics[height=2in, width = 3in]{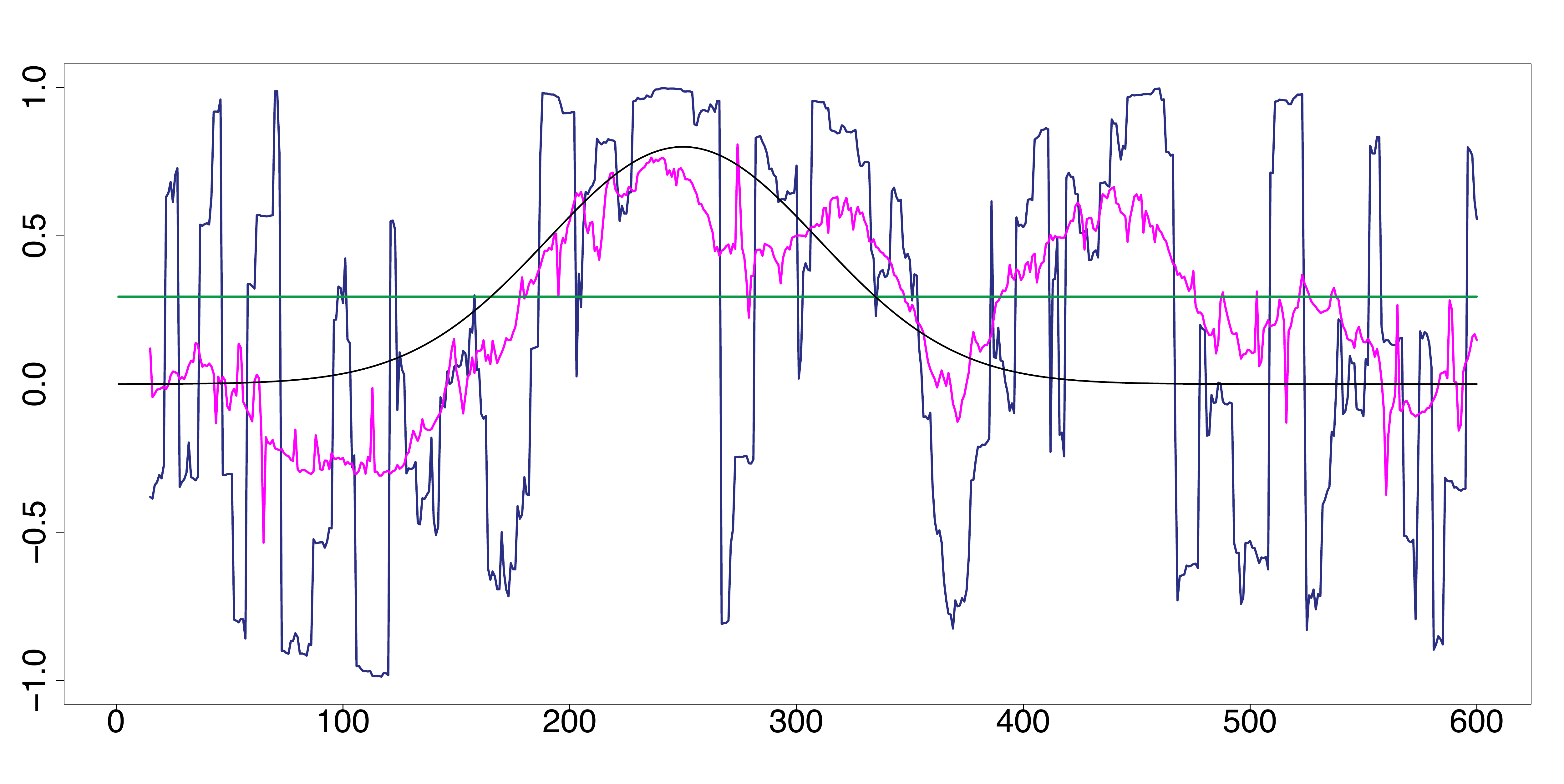}} \\
\caption{Results for single iteration from simulations design D3.
Top row: Underlying pair of time series from bivariate Normal.
Bottom row: Underlying pair of time series from bivariate Cauchy.
Left column: Design 3a (that is, $k=3$). Right column: Design 3b ($k=4$).
The underlying true dynamic correlations are plotted as the black curve.
Blue (SW), magenta (WVGA), green solid (DCC with \textit{rmgarch})
and green dashed (DCC with \textit{rmgarch}) represents the estimates
of dynamic correlation. }
\end{center}
\end{figure}

\begin{figure}[H]
\begin{center}
\subfloat[]{\includegraphics[height=2in, width = 2in]{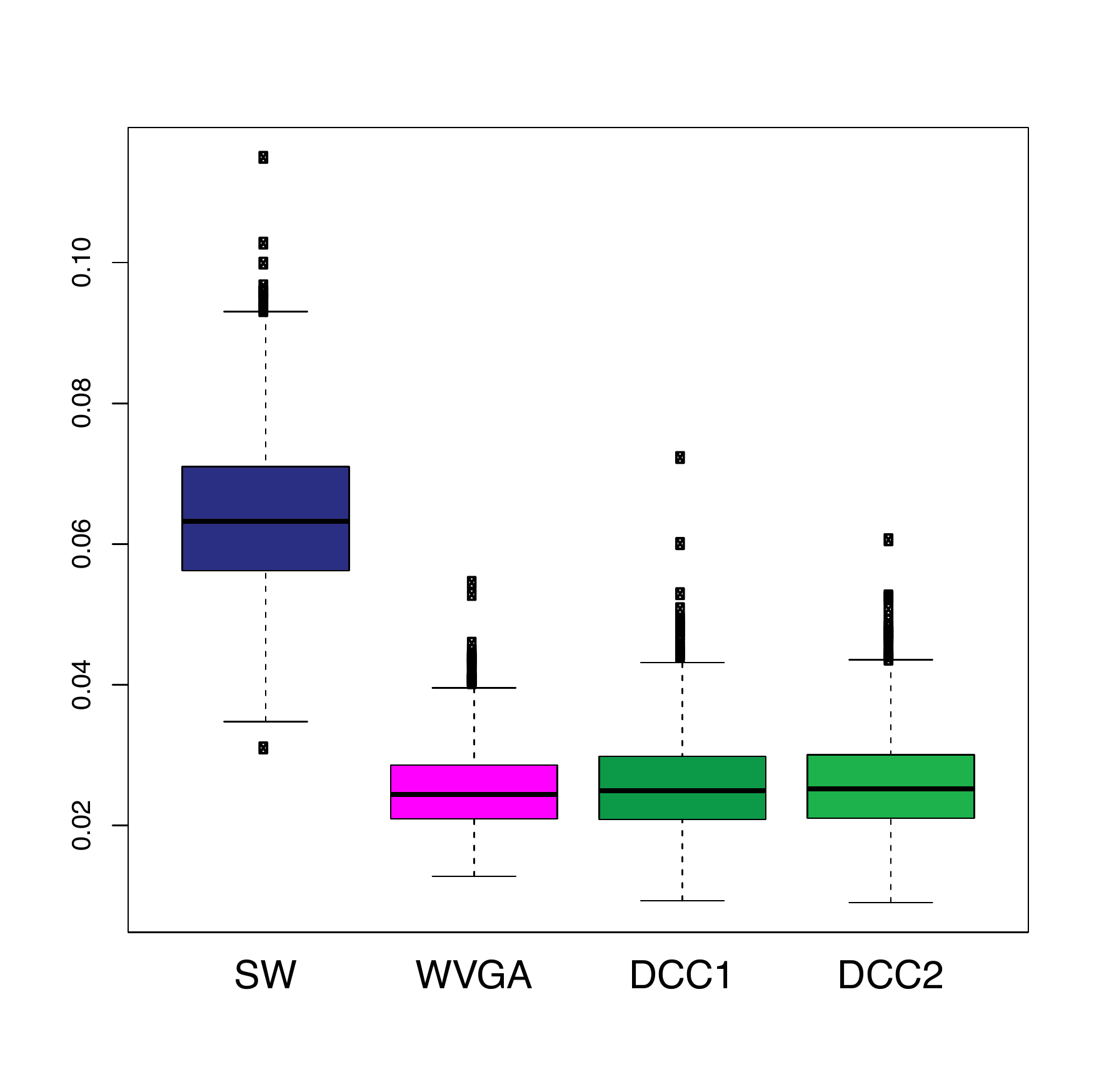}}
\subfloat[]{\includegraphics[height=2in, width = 2in]{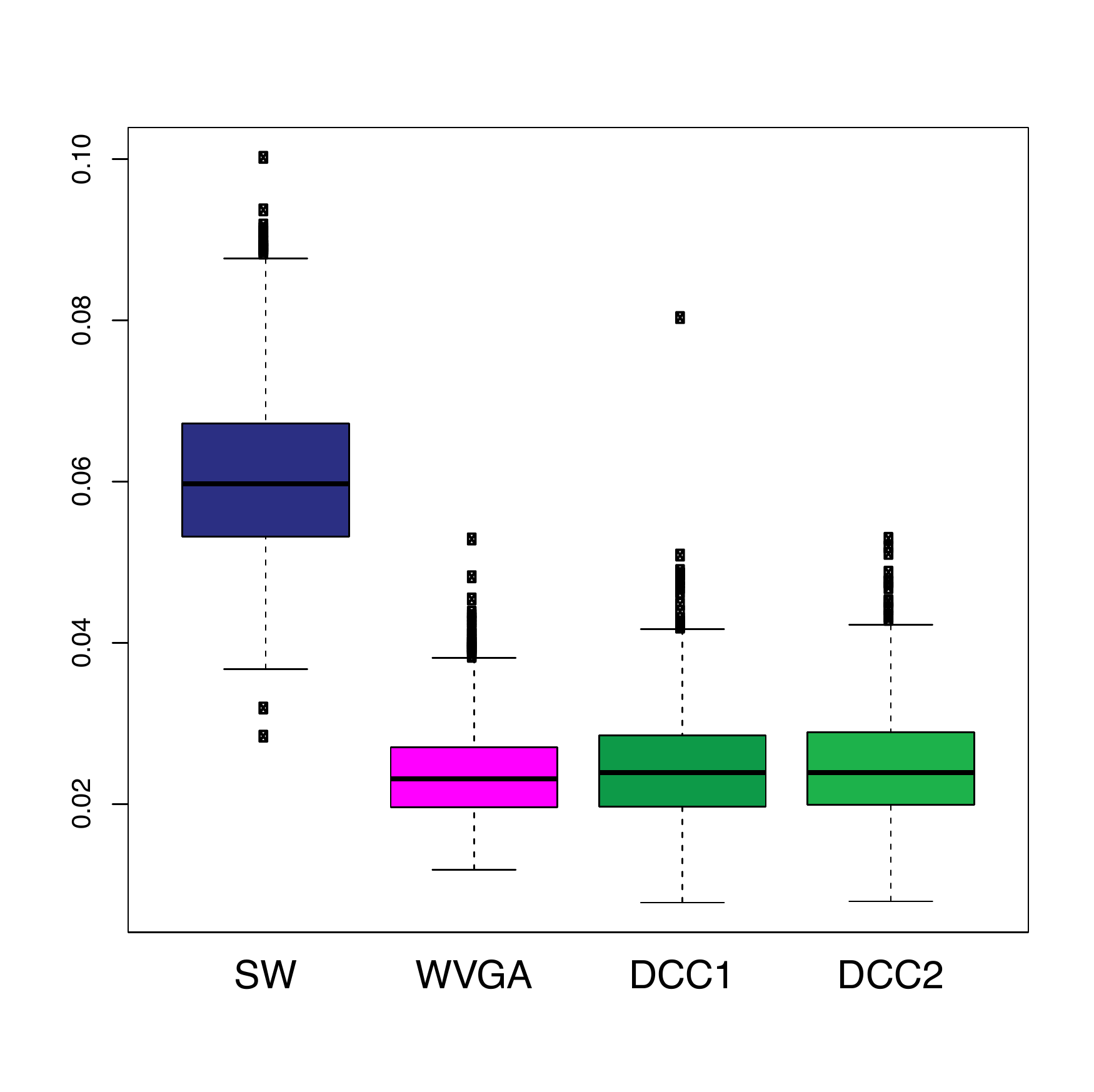}}\\
\subfloat[]{\includegraphics[height=2in, width = 2in]{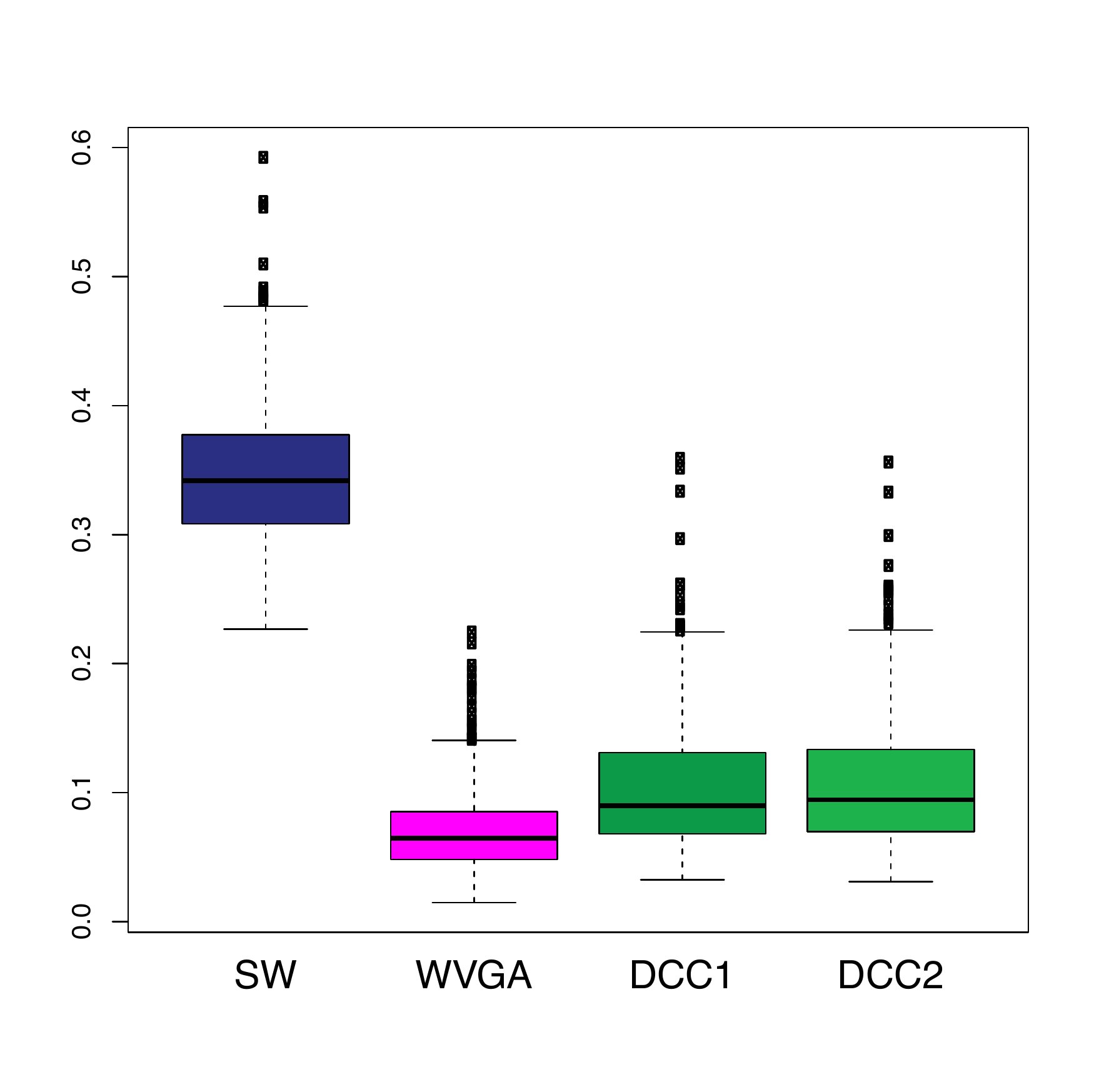}}
\subfloat[]{\includegraphics[height=2in, width = 2in]{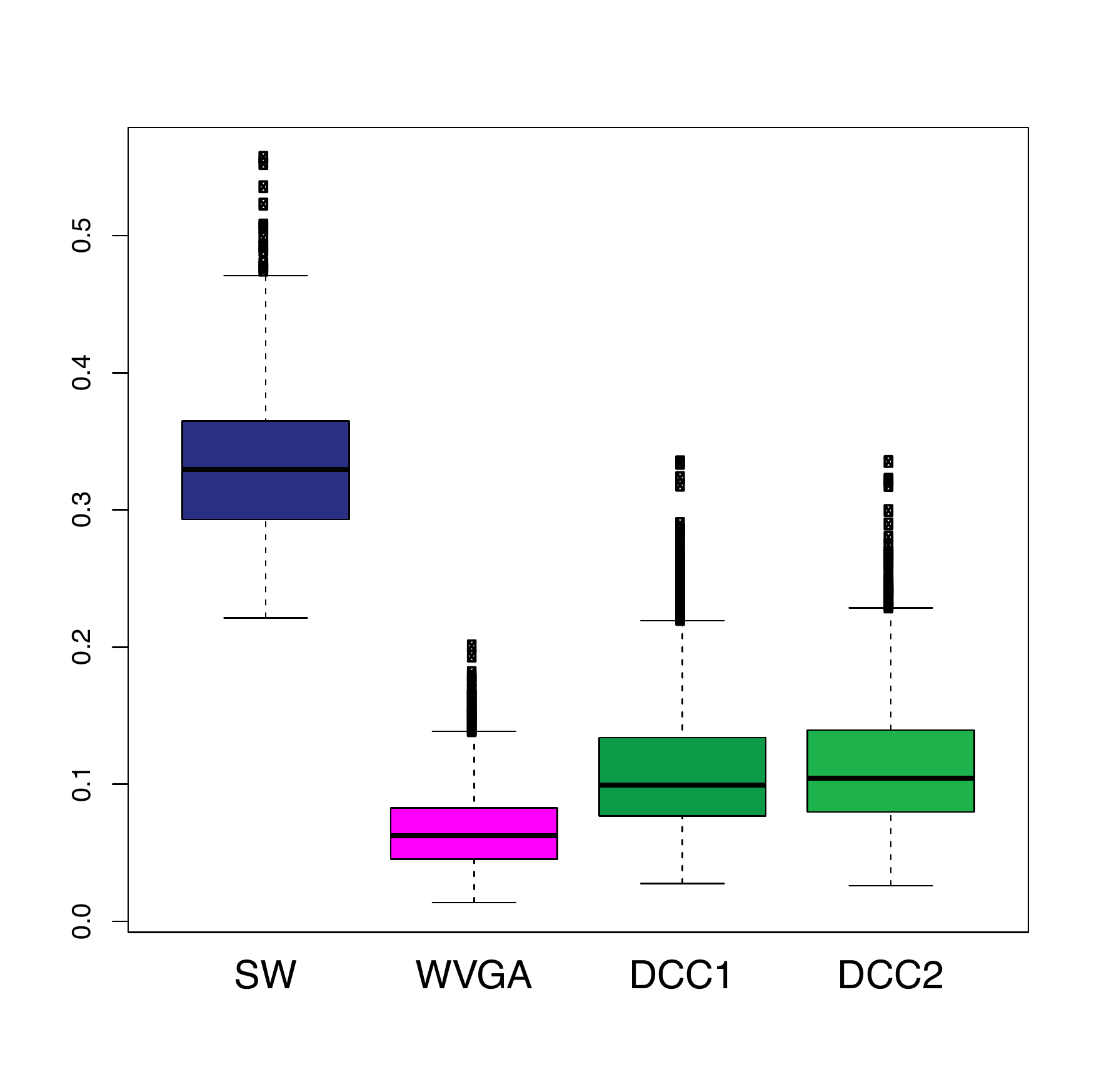}}
\caption{Boxplots of mean squared error of the estimation methods based
on 1000 iterations from simulations design D3. The panels correspond to
bivariate normal (top row), bivariate Cauchy (bottom row), design D3a (left
column) and design D3b (right column), as in figure 5.}
\end{center}
\end{figure}

\subsection*{Stroop task data}

The six times series obtained from six different brain regions of a single subject are shown in Figure 7. Since these data plots did not show the
presence of any extreme values, we would expect the estimates from DCC to be our most accurate estimate of the true underlying correlations. The
results of this data analysis were useful in illustrating the non-convergence problem related to DCC.

\begin{figure}[H]
\begin{center}
\includegraphics[height=3in,width=7in,angle=0]{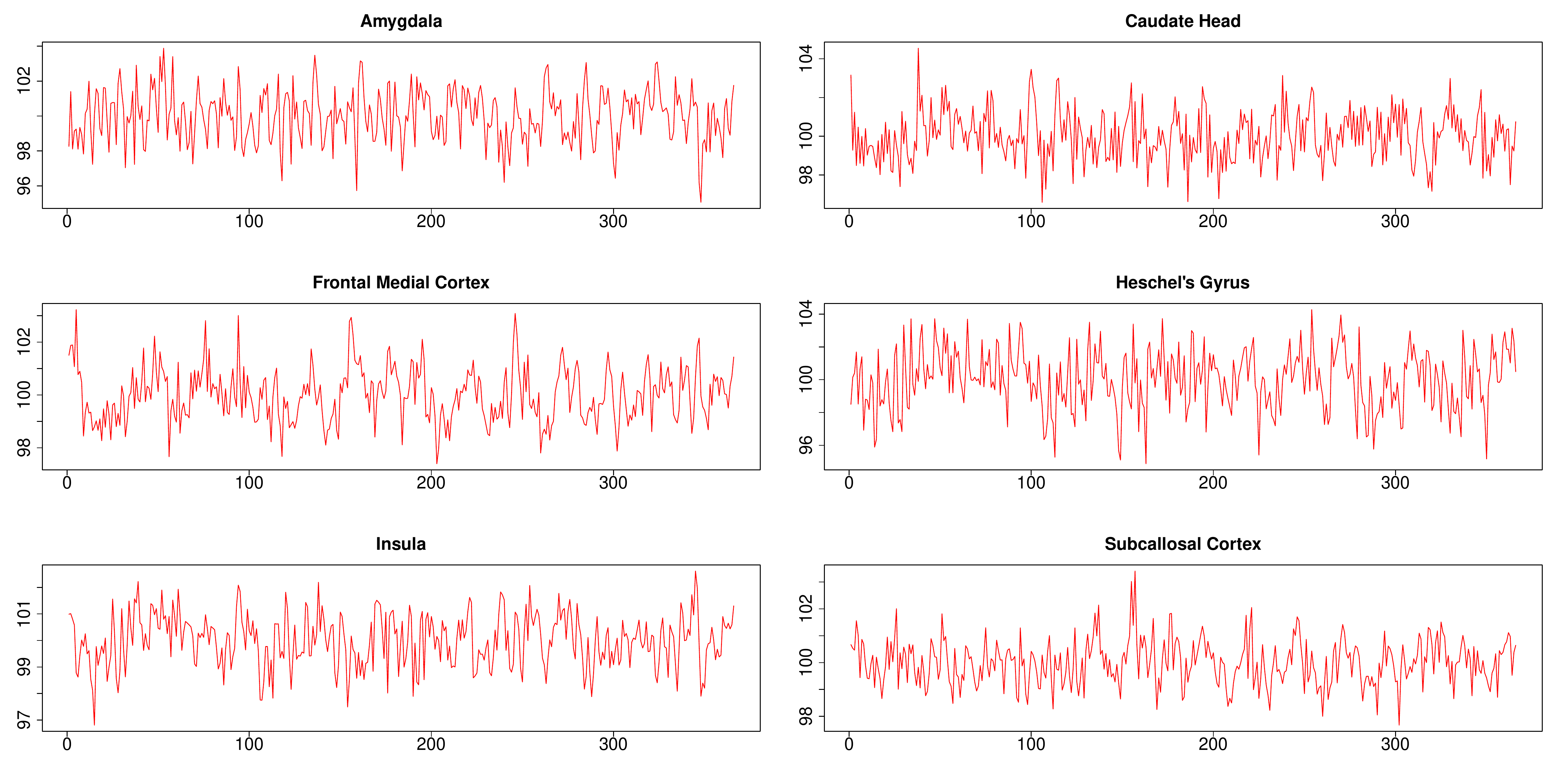}
\caption{Time series from six regions in Stroop task fMRI data}
\end{center}
\end{figure}

Pairwise dynamic correlation plots are shown in Figure 8. Based on the conclusions from our simulation study, the results of the DCC estimation are
likely to be the best estimate of the true dynamic correlation. The results partially validate this expectation. On average, estimates based on SW,
DCC obtained by using the \textit{rmgarch} package in R, and WVGA (blue, solid green, and magenta curves, respectively) indicate roughly similar
values, although the DCC estimate is more or less constant across time while the SW and WVGA estimates fluctuate over time. However, the DCC
estimates obtained using the \textit{ccgarch} R package(dotted green lines in Figure 8) were markedly different from the results of the same analysis
based on the \textit{rmgarch} package. For all pairwise comparisons, the cc-garch based DCC estimates were 1 or very close to 1. These high
correlation values could be caused by the optimization routine used for ML estimation in \textit{ccgarch}, suggesting a lack of convergence problem.
We got similar high values when we tried various other sets of initial values. Conversely, there were other instances (e.g.
in our simulation study) where \textit{ccgarch} provided good (i.e. near correct) estimates, but DCC estimation based on \textit{rmgarch} did not
converge at all. Thus, the issue illustrated by the analysis of the fMRI data is not specific to the \textit{ccgarch} R package per se, but a problem
with DCC in general. This method is dependent on optimization procedures for ML estimation, which could be its Achilles' heel: as with other
optimization routines, occasionally there could be convergence related issues. Convergence problems do not exist for the nonparametric WVGA method
proposed in this paper because it does not involve optimization at any point.

\begin{figure}[H]
\begin{center}
\includegraphics[height=7in,width=7in,angle=0]{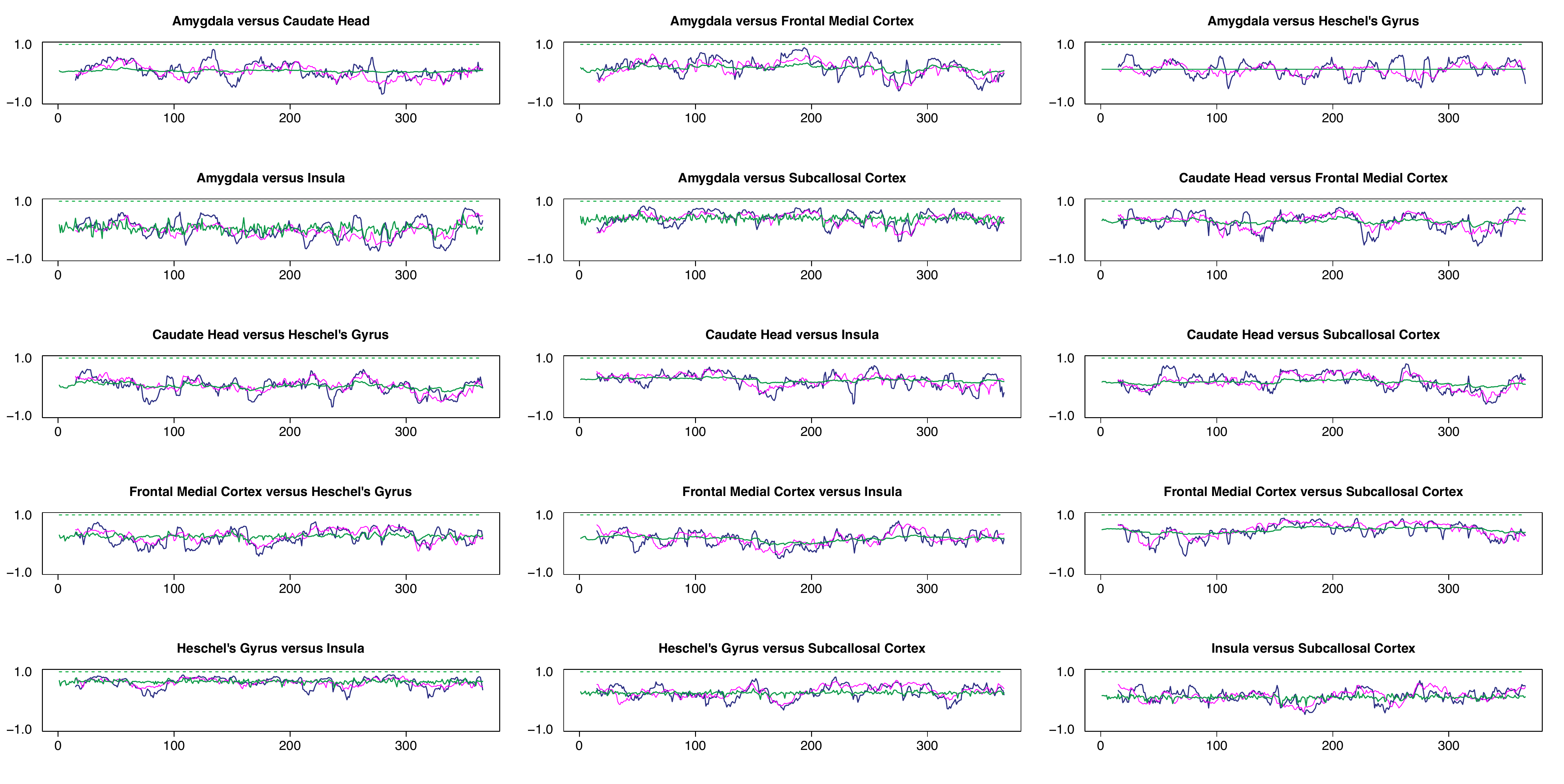}
\caption{Pairwise dynamic correlation estimates for the Stroop task fMRI data.
Blue (SW), magenta (WVGA), green solid (DCC with \textit{rmgarch})
and green dashed (DCC with \textit{ccgarch}). }
\end{center}
\end{figure}

The data analysis presented in this subsection was mainly to illustrate the performance of the different estimation methods for dynamic correlation.
Since the data presented originated from only one subject, not much can be read, interpreted or generalized from the results. Nevertheless, the
highest correlation occurred between Heschel's gyrus and insula (see Table 3 for the mean absolution correlation across all time points for all
pairwise analyses and Table A1 in the Appendix for the corresponding maximum of absolute values of the correlations). Heschel's gyrus contains the
primary auditory cortex (PAC), the first cortical structure to process incoming auditory information [23]. In turn, PAC is in close proximity to the posterior insular cortex [23], to which it is also directly connected [24]. If we use the estimate of dynamic correlation as a proxy measure of
functional connectivity, the data would imply that the Heschel's gyrus and insula are functionally interconnected. This conclusion has clinical
relevance because reduction in the functional connectivity between PAC and insula has been linked to prosody dysfunction in patients
with schizophrenia [25].

\begin{center}
  \tabcolsep=0.11cm
    \begin{tabular}{ | l || c | c | c | c | }                             \hline
        \multicolumn{5}{|c|}{Table 3. Results from Stroop data analysis}   \\ \hline
        \multicolumn{5}{|c|}{Mean of absolute value of correlations across time}   \\ \hline
                                                             & SW         & WVGA       & DCC1             & DCC2    \\ \hline
              Amygdala vs. Caudate Head                      &  $0.075$   & $0.031$    & $1.000$  & $0.069$    \\ \hline
              Amygdala vs. Frontal Medial Cortex             &  $0.210$   & $0.173$    & $1.000$  & $0.169$    \\ \hline
              Amygdala vs. Heschel's Gyrus                   &  $0.138$   & $0.117$    & $1.000$  & $0.129$    \\ \hline
              Amygdala vs. Insula                            &  $0.022$   & $-0.020$   & $1.000$  & $0.047$    \\ \hline
              Amygdala vs. Subcallosal Cortex                &  $0.417$   & $0.378$    & $1.000$  & $0.394$    \\ \hline
              Caudate Head vs. Frontal Medial Cortex         &  $0.280$   & $0.301$    & $1.000$  & $0.299$    \\ \hline
              Caudate Head vs. Heschel's Gyrus               &  $0.010$   & $0.039$    & $1.000$  & $0.025$    \\ \hline
              Caudate Head vs. Insula                        &  $0.214$   & $0.223$    & $1.000$  & $0.240$    \\ \hline
              Caudate Head vs Subcallosal Cortex             &  $0.174$   & $0.155$    & $1.000$  & $0.149$    \\ \hline
              Frontal Medial Cortex vs. Heschel's Gyrus      &  $0.219$   & $0.275$    & $1.000$  & $0.254$    \\ \hline
              Frontal Medial Cortex vs. Insula               &  $0.151$   & $0.179$    & $1.000$  & $0.165$    \\ \hline
              Frontal Medial Cortex vs. Subcallosal Cortex   &  $0.440$   & $0.501$    & $1.000$  & $0.474$    \\ \hline
              Heschel's Gyrus vs. Insula                     &  $0.619$   & $0.622$    & $1.000$  & $0.656$    \\ \hline
              Heschel's Gyrus vs. Subcallosal Cortex         &  $0.274$   & $0.313$    & $1.000$  & $0.261$    \\ \hline
              Insula versus Subcallosal Cortex               &  $0.142$   & $0.142$    & $1.000$  & $0.105$    \\ \hline
    \end{tabular}
\end{center}

\subsection*{LFP data analysis}

Time series corresponding to $~1$ second (1024 data points) from the three tetrodes and one single wire electrode are plotted in Appendix Figure A3.
Tetrodes 1, 2 and 3 were placed in the CA1 field of the hippocampus, while electrode 4 was located in the medial dorsal striatum. All hippocampal
tetrodes recorded a theta oscillation (7-14Hz) typical for this brain area. Signals on tetrodes 2 and 3 were most similar, and included, along with
theta, oscillations of much higher frequency; these oscillations were missing in the signal recorded on tetrode 1. The signal on electrode 4,
recorded in the medial dorsal striatum, lacked the theta rhythm and was visibly different from the other three LFP's. However, in all four recordings a large jump (that is, an extreme value) corresponding to a movement artifact occurred approximately near the 300th time point. The configurations of the four signals were used
to evaluate the performance of SW, DCC, and nonparametric WVGA methods discussed above. To estimate the dynamic correlations between the signals in
the presence of extreme values, we cut out segments of data between time points 200 and 400 for each of the four time series (left panels of Figure
9). We expected to obtain high correlations between all segments around the 300th data point, where the movement artifact registered in all four
recordings. Outside of this point, we expected high correlation between LFPs 2 and 3, somewhat lower correlation between LFP 1 and either LFP 2 or
LFP 3, and low correlations between LFP 4 and the signals on any of the three tetrodes. Furthermore, because of the presence of extreme values, we
also expected that our nonparametric WVGA would generate the best estimate.

Second, we also cut segments between time points 415 and 600 (plotted in the right panels in Figure 9). Visual inspection suggested a change in
signal configuration around the 50th point of these fragments which should be reflected in changes in the dynamic correlations among LFPs 1-3 (left
vs. the right of the 50th time point in Figure 9, right panels). At the same time, we expected little correlation between any of the LFPs 1-3 on one
hand and LFP 4 on the other hand.

\begin{figure}[H]
\begin{center}
\includegraphics[height=3in,width=7in,angle=0]{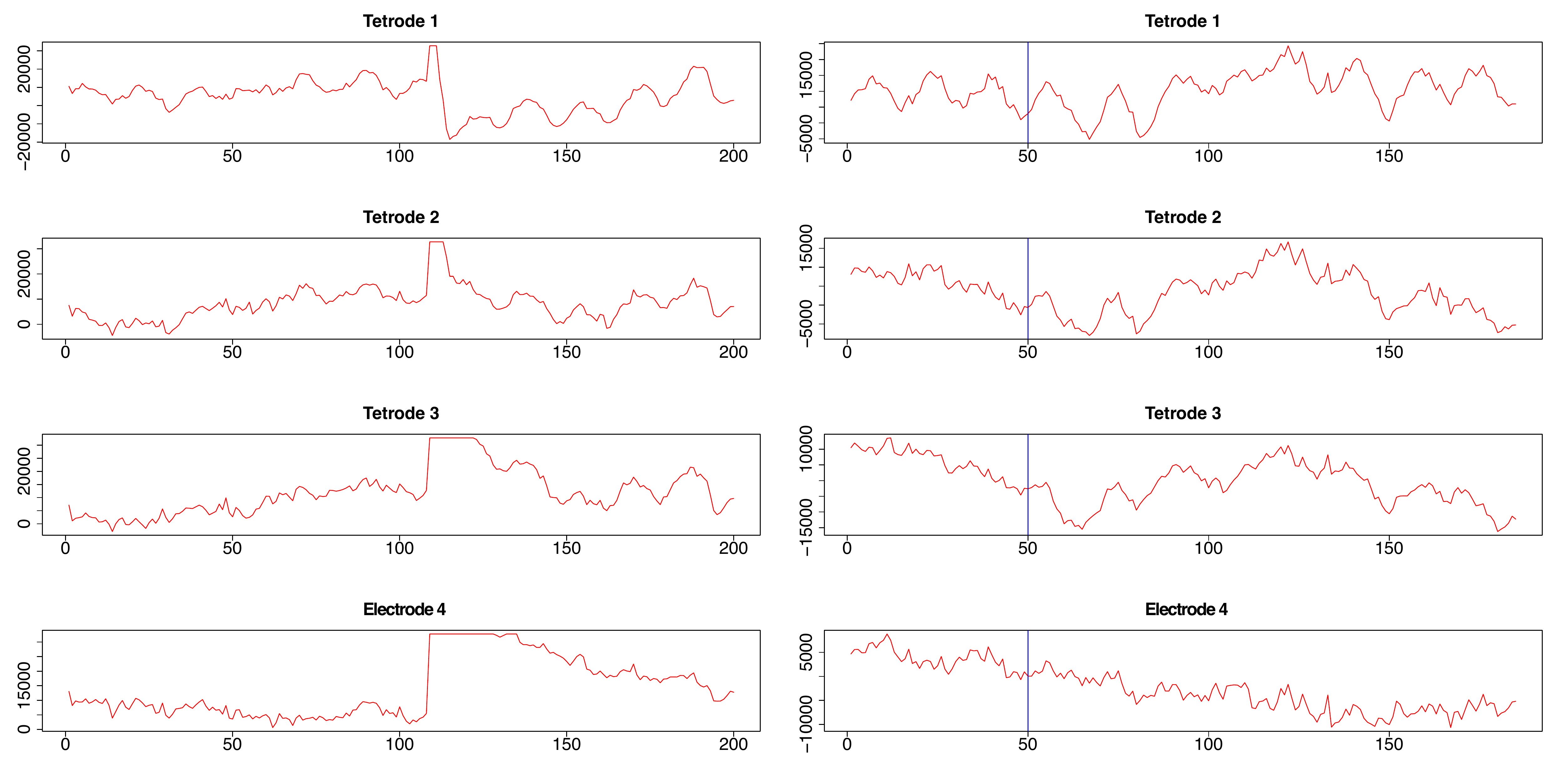}
\caption{Segments of LFP time series from four electrodes implanted in the CA1 field
of the hippocampus (tetrodes 1-3) and medial dorsal striatum (electrode 4) in the brain of
the same rat. Left column: Segments between time points 200 and 400 from the respective time
series plotted in appendix figure A3. Right column: segments between time points 415 and 600 of
time series seen in figure A3. $50^{th}$ time point within these segments is marked by a blue vertical line.}
\end{center}
\end{figure}

The results from pairwise analyses of dynamic correlations for the segments in the left side of figure 9 are plotted in figure 10. In all these
plots, the estimates generated by the SW and DCC methods fluctuated rapidly compared to nonparametric WVGA, suggesting that the presence of extreme values affected SW and DCC much more than WVGA. As in the fMRI analysis, the two DCC estimates based on the two different R
packages differed markedly. Moreover, the SW method did not always result in a correlation value (see discontinuities in the blue curve in Figure
10) because in some portions the original signals consisted of series of equal numbers with zero standard deviation. Thus, the WVGA method generated
the best estimates for this set of time series.

Visual comparison of LFPs 1 and 3 (first and third right panels in Figure 9, top left panel in Figure A4) indicated that although both signals were
dominated by the theta rhythm and showed a movement artifact, they were overall dissimilar, particularly at the later time points. The WVGA estimates (top right in Figure 10, top right in Figure A4) reflected accurately these variations, showing low correlation at the beginning of the signals
that raised to a maximum correlation at the point of the extreme values corresponding to the movement artifact, dipped to negative values in the
immediate aftermath of the peaks, and returned to positive values at the very end. Similarly, visual inspection of the tetrodes 2 and 3 signals
(middle two panels in the left side of Figure 9; middle left panel in Figure A4), suggested consistent high correlations across all time points
within the segments. The results of the nonparametric WVGA analysis showed that this was indeed the case (magenta curve in the middle right panel of
Figure 10; middle right panels of Figure A4). There was no significant dynamic shift for the magenta curve and its values were above 0.5 for a significant
portion of the segment. In contrast, the other estimates fluctuated rapidly (blue and green curves, middle right panel in Figure 10) which neither
makes sense biologically, nor verifies our understanding based on visual inspection of the original pair of LFPs. Finally, the WVGA estimate
indicated that the dynamic correlation between any of the hippocampal signals (LFPs 1-3) and the striatal signal (LFP4) was consistently low except
in the section of extreme values, where, not surprisingly, the values were high (middle and bottom right, bottom left in Figure 10, bottom right in
figure A4). This assessment corresponded with our expectation that the LFPs originating in the two different brain areas would be uncorrelated.

\begin{figure}[H]
\begin{center}
\includegraphics[height=3in,width=7in,angle=0]{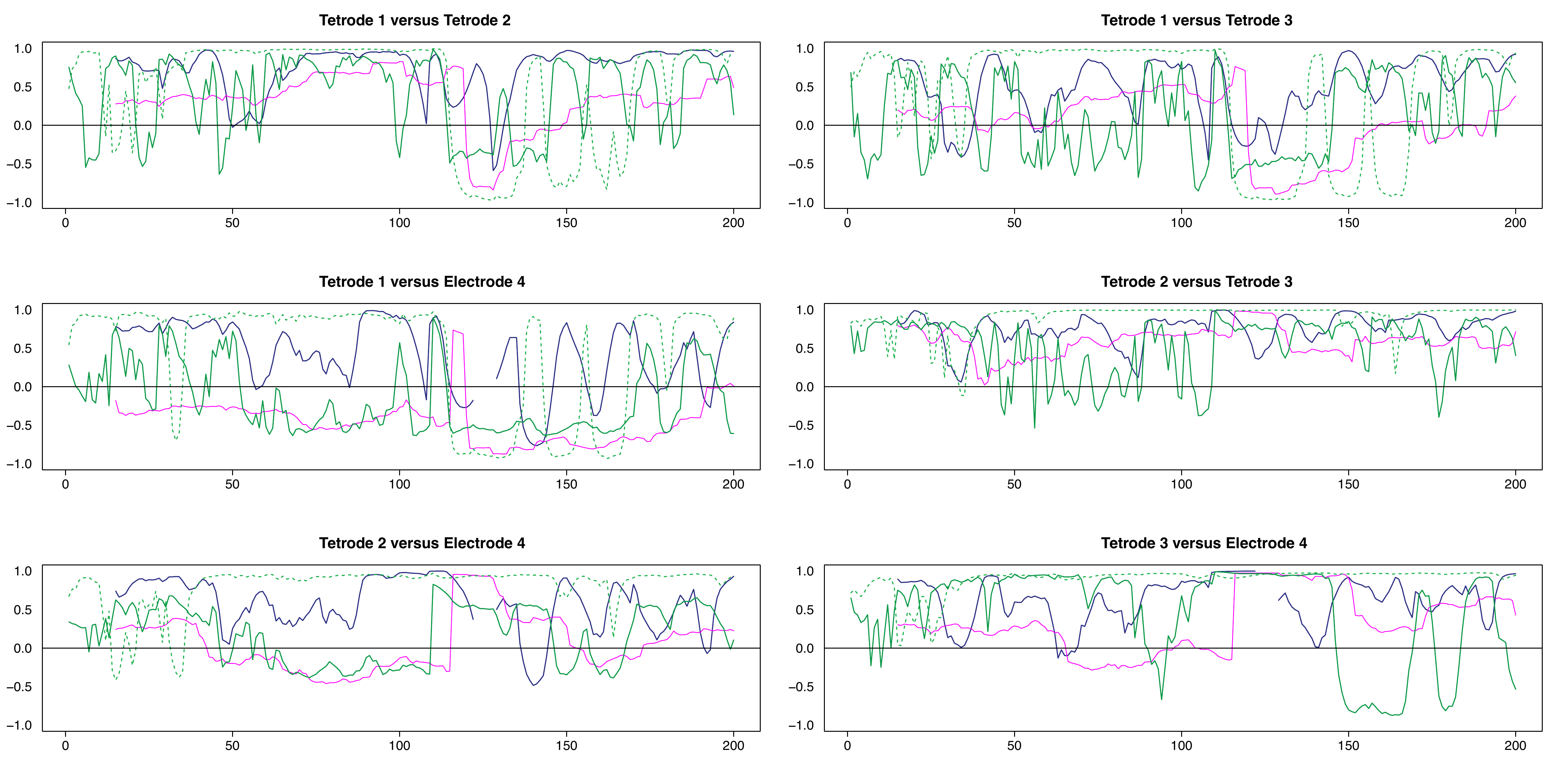}
\caption{Pairwise dynamic correlations for time series plotted in the left column panels
in figure 9. Blue (SW), green solid (DCC with \textit{rmgarch}),
green dashed (DCC with \textit{ccgarch}), and magenta (WVGA).}
\end{center}
\end{figure}

The results from pairwise analyses of dynamic correlations for the segments in the right side of Figure 9 are plotted in Figures 11 and A5. As
previously, the estimates based on SW and DCC fluctuated rapidly and the two DCC-based estimates differed significantly in all cases, even for the
LFPs 2 and 3, which were very similar. In contrast, WVGA analysis reflected correctly the switch in dynamic correlation before and after the 50th
time point and the higher similarity between the signals on tetrodes 2 and 3 vs. signal on tetrode 1 (top left panel and top two right panels in
Figure 11, top two rows in Figure A5). Except for the segment before the 50th point, the nonparametric WVGA estimate of the dynamic correlation
between tetrodes 2 and 3 was consistently above 0.5, confirming our visual evaluation. The estimate for the same time points of the dynamic
correlations of electrode 1 with electrodes 2 and 3, respectively, showed near zero values in the beginning that gradually rose and after the
$50^{th}$ time point, the values were mostly above 0.5. These results again confirmed our visual observation of the original time series. Finally, as
expected, the nonparametric WVGA estimates of dynamic correlations between hippocampal and striatal LFPs were near zero for most of the time points.
Thus, the nonparametric WVGA method led to meaningful results.

\begin{figure}[H]
\begin{center}
\includegraphics[height=3in,width=7in,angle=0]{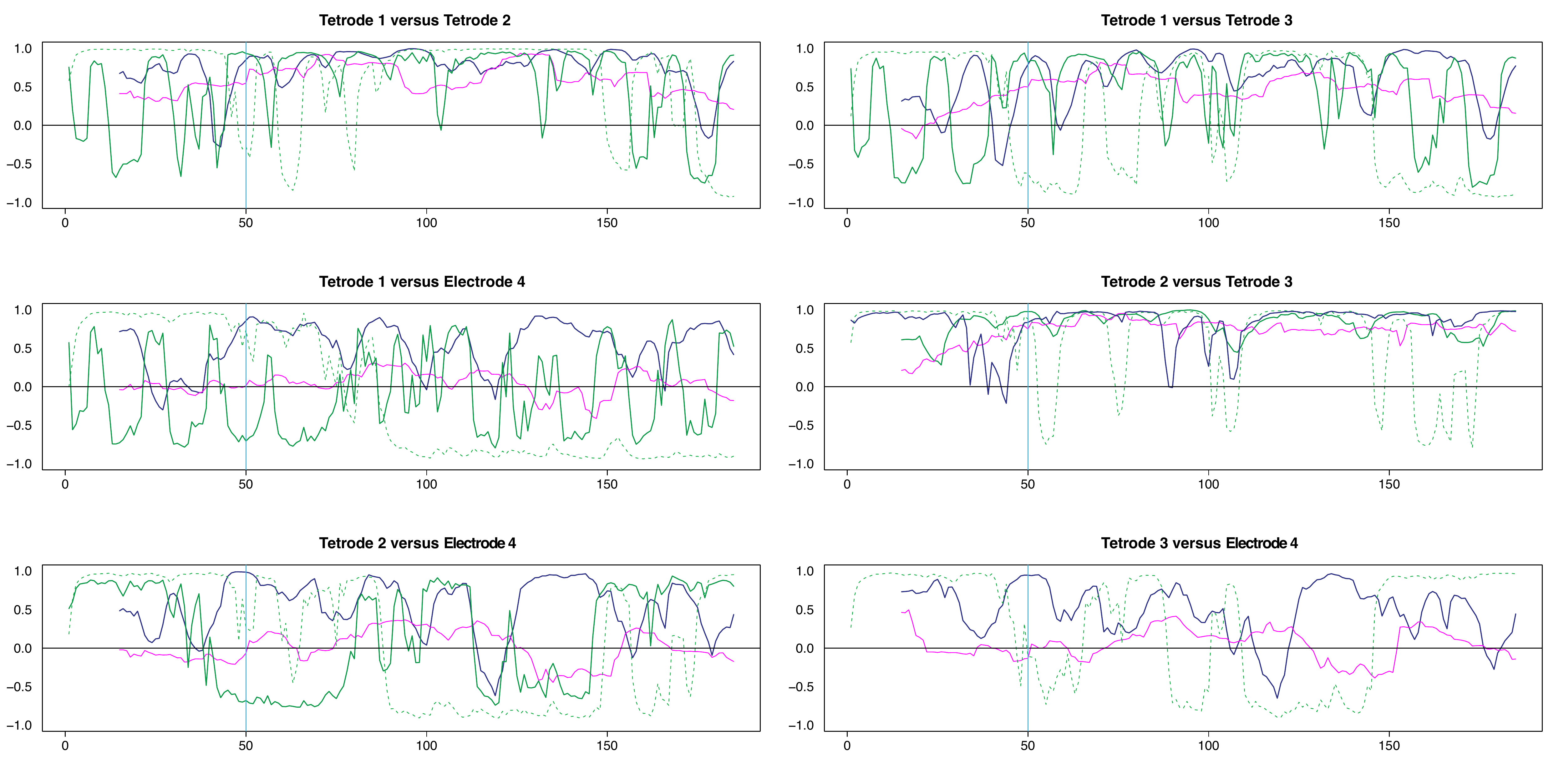}
\caption{Pairwise dynamic correlations for time series plotted in the right column panels in figure 9. Blue (SW), magenta (WVGA), green solid (DCC with \textit{rmgarch}) and green dashed (DCC with \textit{ccgarch}). $50^{th}$ time point within these segments is marked by a light blue vertical line. The green dashed curve in the lower right panel is missing, indicating a convergence problem with the DCC with \textit{ccgarch} in this case.}
\end{center}
\end{figure}

\section*{Discussion and Conclusions}

Understanding the dynamic correlation underlying a pair of time series is of interest in many areas including neuroscience. Currently,
existing techniques such as sliding window (SW) or GARCH-based DCC can be applied to this estimation problem. Linquist and co-authors[8] evaluated
the existing methods and concluded that DCC is an attractive option for effective dynamic correlation estimation in fMRI data. Our analysis confirmed that DCC performs well for nicely behaved time series from a bivariate normal distribution. However, for time series with extreme values, common in
neuroscience, DCC performs much more poorly. To address this problem, we propose a nonparametric approach to the estimation of dynamic correlation by
adapting the weighted visibility graph algorithm, a relatively novel method introduced by physicists [11,16, 17, 26] that converts a time series
into a graph. This method is independent from finite moment assumptions, rendering it robust to extreme values.

To study how the nonparametric WVGA compares with the SW and DCC methods, we conducted extensive simulations as in Lindquist
\textit{et al} [8]. We also compared the results obtained by applying the three methods to one real fMRI data set and to one LFP data set with known
properties and extreme values recorded in an awake rat. These analyses indicated that the WVGA based method led to more biologically meaningful
results and performed better in the presence of outliers than SW or DCC.

We also found that both in the simulations and in the analysis of the fMRI data set, DCC exhibited convergence problems. We speculate that these
issues are possibly generated by the optimization algorithms involved. The state-of-the-art method for estimating the parameters in the DCC model
uses a 2-stage ML method involving numerical optimization, a procedure whose routines may exhibit convergence issues - non-convergence, or
convergence to the wrong solutions. Convergence problems do not exist for  WVGA based method because it does not involve the estimation of any
parameters.

The nonparametric WVGA method performs very well in the scenarios for which it was intended to be utilized, but a deeper understanding of this
performance remains to be yet developed.  Our adaptation of WVGA is a bit ad-hoc in nature, unlike, for example, DCC, which is based on
well-established statistical frameworks such as GARCH and quasi-likelihood estimation. One factor that may contribute to the optimal performance of
the nonparametric WVGA method could be the combination of local and global aspects of its algorithm. Method $\#1$ (SW) is based on only local
information - correlation at each time point is based only on the values within the sliding window. As mentioned in Lindquist \textit{et al} [8], in
these conditions the dynamic profiles of even random noise may show compelling changes in correlation across time. The local nature of SW could be
transformed into a more global nature by increasing the window-size, but increasing the window-size beyond a particular level will lead to the method
overlooking real dynamic changes. Method $\#2$ is much more of a global estimation approach. In DCC, only the parameters of the
covariance/correlation matrix are considered to vary while all other parameters ($\omega_{1}, \alpha_{1}, \beta_{1}, \omega_{2}, \alpha_{2},
\beta_{2}$ in Lindquist \textit{et al}'s notation) remain constant. The new approach based on WVGA incorporates both local and global components.
Since it is window-based, it is partly local in nature like SW. However, for time points within each window the correlations calculated involve
median weight vectors with similar length as that of the original time series and consisting of weights from all time points from the entire series.
This approach confers global character to the new method. The combined local and global aspects of the nonparametric WVGA method may result in
optimal performance on data sets that do not conform to normality and contain extreme values. While this aspect remains to be explained in more
detail through further work, it is clear that extreme values do not pose a problem for our nonparametric WVGA, which is based on a weighted graph
where the weights are calculated using an \textit{arctan} function. \textit{arctan}, like other functions used in robust estimation such as Huber's function, trimming function or Tukey's bisquare function reduces the influence of outlying values by applying a cap (see section 14.5 in [27]). In contrast, DCC, like most other parametric methods, is influenced disproportionately by the extreme values.

Although the ad hoc nature of our adaptation of WVGA
is not a limitation per se because the method performs well in practice, gaining a
deeper understanding of why it works would facilitate improving the current
version presented in this paper. Visibility graphs other than the one utilized in this paper
have been discussed in the literature, the most prominent among them being the horizontal visibility graph.
Future work will include exploring the performance of adaptations of these alternate visibility graph algorithms
for estimating dynamic correlation.

\section*{Acknowledgements}

Aparna John would like to acknowledge Open Cloud Endowment Fellowship from the Open Cloud Institute at UTSA. Toshikazu Ikuta would like to acknowledge Brain and Behavior Research Foundations's NARSAD Young Investigator Award. Janina Ferbinteanu's research was supported by National Institute of Mental Health (NIMH) grant R21MH106708. Majnu John's work was supported in part by grants from the NIMH for an Advanced Center for Intervention and Services Research (P30 MH090590) and a Center for Intervention Development and Applied Research (P50 MH080173), and also in part by the following NIMH grants: R01MH109508 (PI: Anil K Malhotra) and R01MH108654 (PI: Anil K Malhotra).

\section*{Appendix} \label{App:AppendixA}

\renewcommand{\thefigure}{A\arabic{figure}}
\setcounter{figure}{0}

\begin{figure}[H]
\begin{center}
\includegraphics[height=3in,width=7in,angle=0]{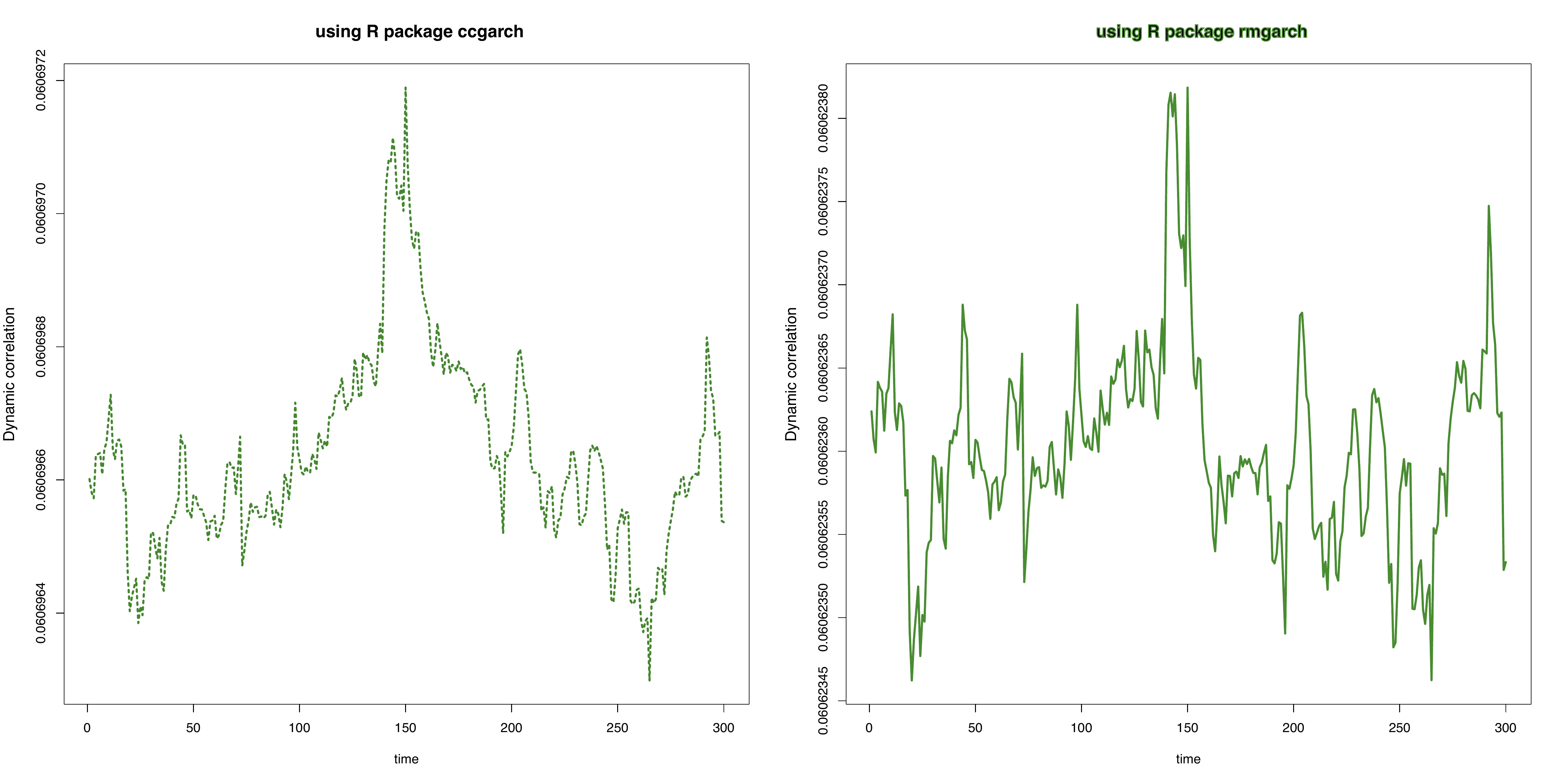}
\caption{The green curves (DCC estimates) in the bottom panel of figure 1,
with $y$-axis zoomed in, to see that the curves are not really straight lines.}
\end{center}
\end{figure}

\begin{figure}[H]
\begin{center}
\includegraphics[height=3in,width=7in,angle=0]{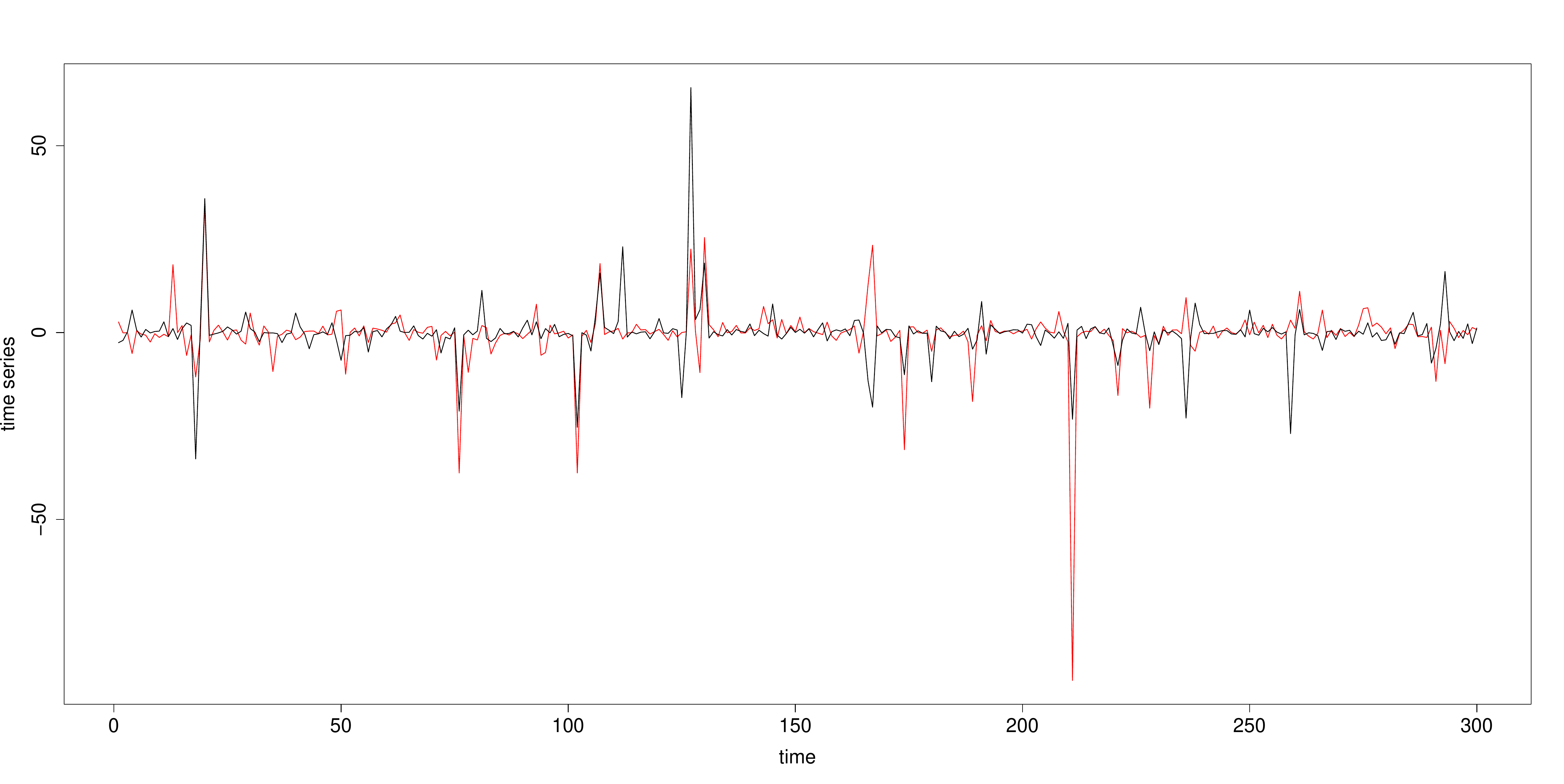}
\caption{The time series plot in the top panel in figure 2, with the $y$-axis
margins expanded.}
\end{center}
\end{figure}

\begin{center}
  \tabcolsep=0.11cm
    \begin{tabular}{ | l || c | c | c | c | }                             \hline
        \multicolumn{5}{|c|}{Table A1. Results from Stroop data analysis}   \\ \hline
         \multicolumn{5}{|c|}{Maximum of absolute value of correlations across time}   \\ \hline
                                                             & SW         & WVGA               & DCC1             & DCC2    \\
                                                             \hline
              Amygdala vs. Caudate Head                      &  $0.823$   & $0.507$    & $1.000$  & $0.194$    \\ \hline
              Amygdala vs. Frontal Medial Cortex             &  $0.882$   & $0.666$    & $1.000$  & $0.347$    \\ \hline
              Amygdala vs. Heschel's Gyrus                   &  $0.657$   & $0.504$    & $1.000$  & $0.129$    \\ \hline
              Amygdala vs. Insula                            &  $0.767$   & $0.584$    & $1.000$  & $0.555$    \\ \hline
              Amygdala vs. Subcallosal Cortex                &  $0.830$   & $0.692$    & $1.000$  & $0.674$    \\ \hline
              Caudate Head vs. Frontal Medial Cortex         &  $0.795$   & $0.701$    & $1.000$  & $0.467$    \\ \hline
              Caudate Head vs. Heschel's Gyrus               &  $0.606$   & $0.492$    & $1.000$  & $0.254$    \\ \hline
              Caudate Head vs. Insula                        &  $0.736$   & $0.602$    & $1.000$  & $0.362$    \\ \hline
              Caudate Head vs Subcallosal Cortex             &  $0.800$   & $0.624$    & $1.000$  & $0.280$    \\ \hline
              Frontal Medial Cortex vs. Heschel's Gyrus      &  $0.757$   & $0.629$    & $1.000$  & $0.461$    \\ \hline
              Frontal Medial Cortex vs. Insula               &  $0.789$   & $0.684$    & $1.000$  & $0.302$    \\ \hline
              Frontal Medial Cortex vs. Subcallosal Cortex   &  $0.886$   & $0.831$    & $1.000$  & $0.597$    \\ \hline
              Heschel's Gyrus vs. Insula                     &  $0.892$   & $0.851$    & $1.000$  & $0.799$    \\ \hline
              Heschel's Gyrus vs. Subcallosal Cortex         &  $0.817$   & $0.703$    & $1.000$  & $0.419$    \\ \hline
              Insula versus Subcallosal Cortex               &  $0.693$   & $0.599$    & $1.000$  & $0.311$    \\ \hline
    \end{tabular}
\end{center}

\begin{figure}[H]
\begin{center}
\includegraphics[height=3in,width=7in,angle=0]{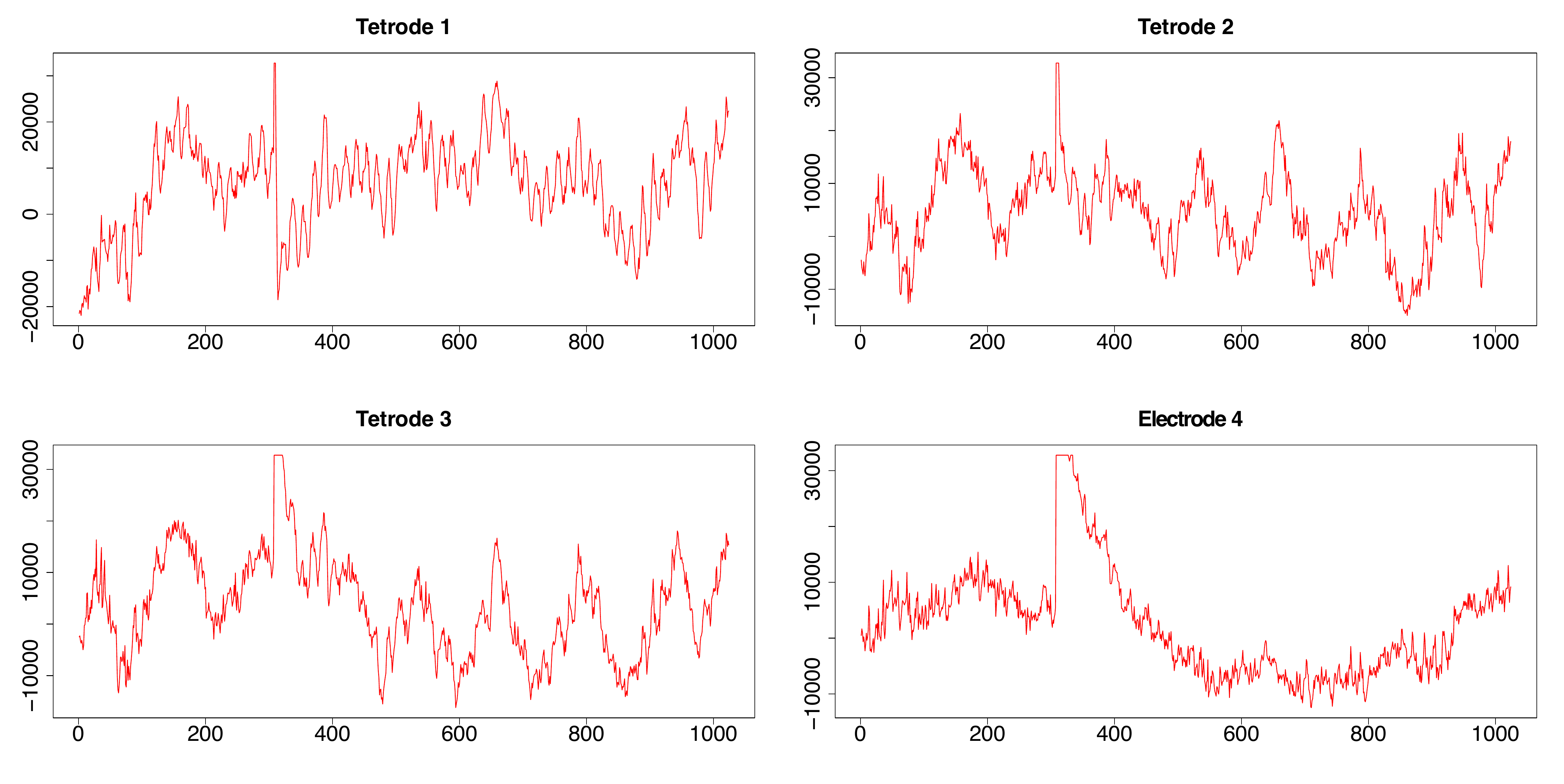}
\caption{LFP time series related to two consecutive time stamps (1024 data points sampled at 1KHz, approximately 1 sec) from four tetrodes.
From each of these time series, segments between time points 200 and 400, and between
415 and 600 were cut out and plotted in left and right panels, respectively, of figure 9.
These segments were analyzed for illustrating the dynamic correlation estimation methods.}
\end{center}
\end{figure}

\begin{figure}[H]
\begin{center}
  \includegraphics[height=3in,width=7in,angle=0]{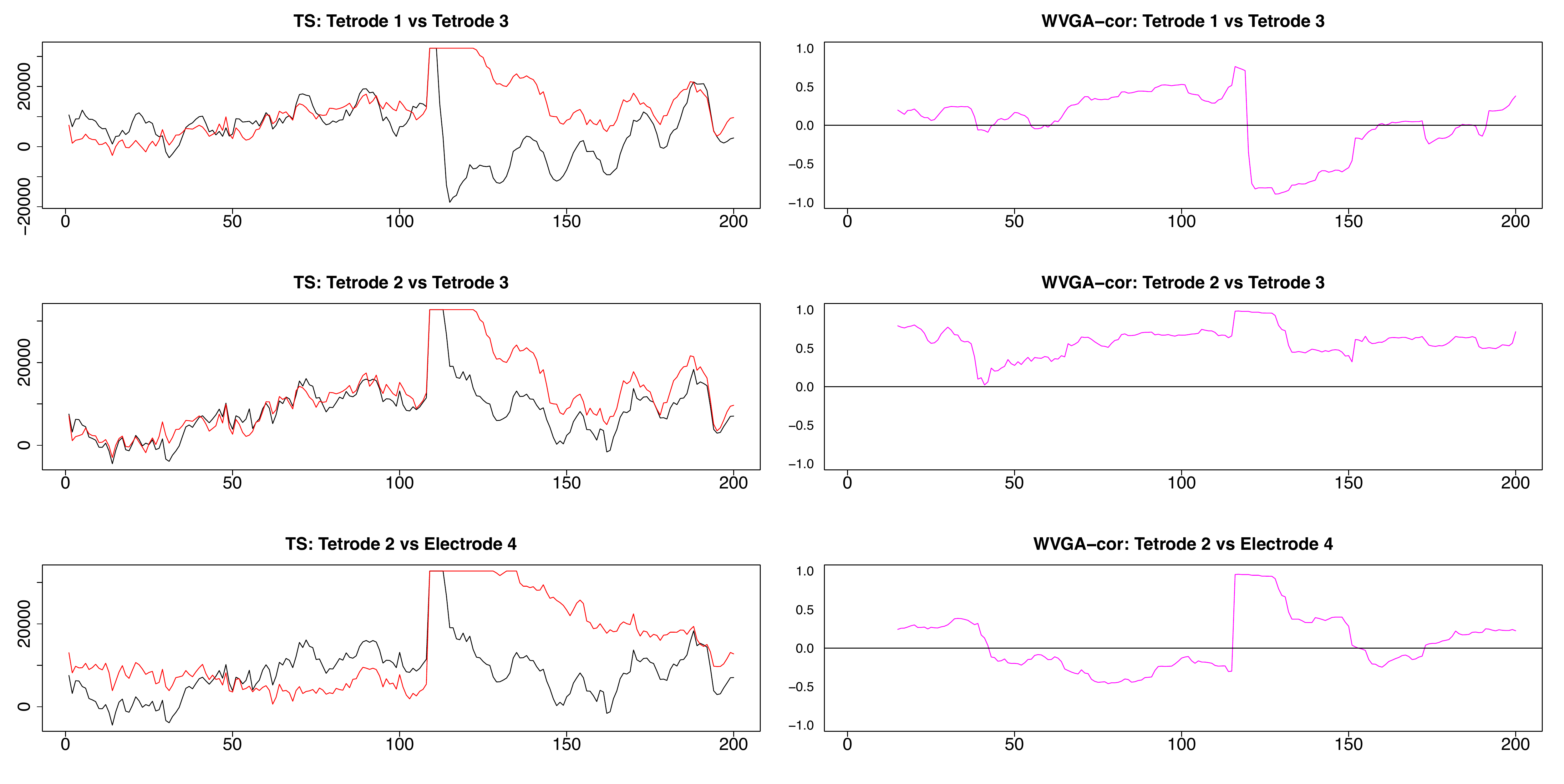}
  \caption{Three plots selected from the pairwise plots in Figure 10. LFP pairs are shown in the left column, the resulting dynamic correlations
  computed by using the nonparametric WVGA method are shown in the right column. In all cases, the maximum correlation occurs at the time point corresponding
  to the stimulus artifact. Overall, the highest correlations resulted for the tetrodes 2 and 3 LFPs which are the most similar, while the dynamic
  correlation between the very different LFPs of tetrode 2 and electrode 4 were dominated by low values.}
\end{center}
\end{figure}

\begin{figure}[H]
\begin{center}
  \includegraphics[height=3in,width=7in,angle=0]{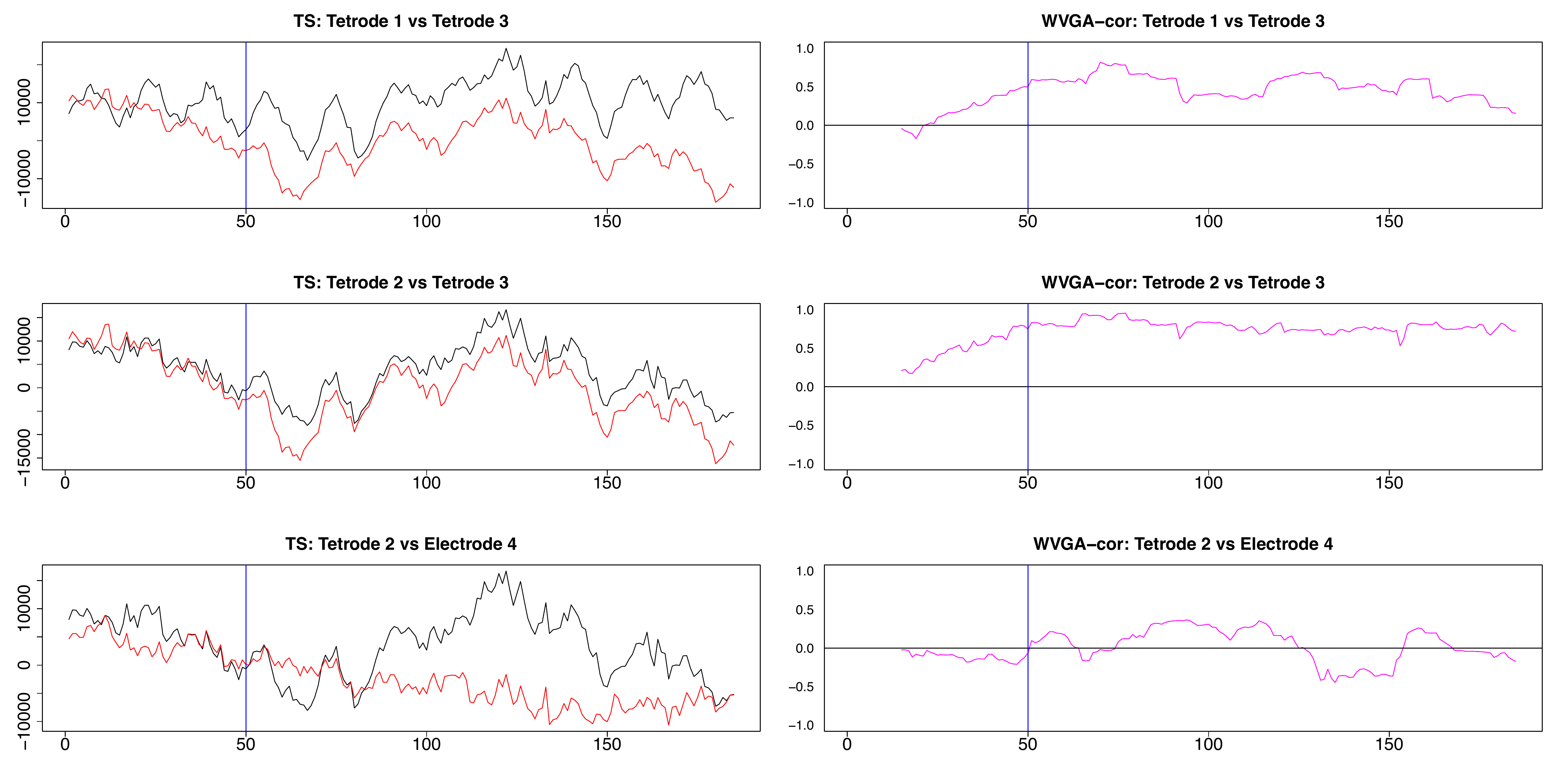}
  \caption{Three plots selected from the pairwise plots in Figure 11. LFP pairs are shown in the left column, the resulting dynamic correlations
  computed by using the nonparametric WVGA method are shown in the right column. The hippocampal LFPs become more similar after the 50th time point and the dynamic correlations in the top two panels reflect adequately this change. The LFP's recorded in hippocampus and medial dorsal striatum are very different, a characteristic reflected in the low values of the corresponding dynamic correlation. }
\end{center}
\end{figure}

\end{document}